\documentclass{iopart}
\usepackage{hyperref}
\usepackage{cite}
\usepackage{graphicx}

\newcommand{\un}[1]{\:\mathrm{#1}}

\begin{document}


\title[Formation of solitary microstructure\dots]{Formation of solitary microstructure and ablation into transparent dielectric by a subnanosecond laser pulse}

\author{V.~A.~Khokhlov$^1$, N.~A.~Inogamov$^{1,2}$, V.~V.~Zhakhovsky$^{2,1}$, Yu.~V.~Petrov$^{1,3}$}
\ead{v\_a\_kh@mail.ru}
\address{$^1$ Landau Institute for Theoretical Physics of RAS, 1-A Akademika Semenova av., Chernogolovka, Moscow Region, 142432, Russia}
\address{$^2$ Dukhov Research Institute of Automatics, 22 Sushchevskaya st., Moscow, 127055, Russia}
\address{$^3$ Moscow Institute of Physics and Technology, 9 Institutskii per., Dolgoprudny, Moscow Region, 141700, Russia}

\begin{abstract}
 Laser ablation in liquid (LAL) is important technique used for formation of nanoparticles (NP).
 The LAL processes cover logarithmically wide range of spatiotemporal scales and is not fully understood.
 The NP produced by LAL are rather expensive, thus optimization of involved processes is valuable.
 As the first step to such optimizations more deep understanding is necessary.
 We employ physical models and computer simulations by thermodynamic, hydrodynamic, and molecular dynamics codes in this direction.
 Absorbing light metal expanding into transparent solid or liquid dielectrics is considered.
 We analyze an interplay between diffusion, hydrodynamic instability, and decrease of surface tension down to zero value caused by strong heating and compression
   transferring matter into state of overcritical fluids.
 The primary NPs appear during expansion and cooling of diffusion zone when pressure in this zone drops below critical pressure for a metal.
 Long evolution from the overcritical states to states below a critical point for a metal and down to critical point of liquid and deeply down to surrounding pressure of 1 bar is followed.
 Conductive heating of liquid from hot metal is significant.
 
\noindent {\bf keywords:} laser ablation, instability, condensation

\end{abstract}




\section{Introduction}


 Paper presents theoretical and simulation results on laser ablation in liquid (LAL) or into initially solid transparent dielectrics;
   gold-water and gold-silica pairs are investigated.
 We consider and compare ultrashort $\tau_L=100$ fs and subnanosecond $\tau_L=50$ ps pulses $I(t)=I_0\exp(-t^2/\tau_L^2)$
  and vary absorbed fluence $F_{abs}=\int_{-\infty}^{\infty} I_{abs}(t) \, dt$ in a wide range up to values limited by optical breakdown of dielectrics;
   threshold values for breakdown are $I_{inc}\sim 10^{11}$ W/cm$\!^2$ for the subnanosecond case with $h\nu\approx 1.2$ eV \cite{Rubenchik:opt.breakdown:PRL:1995,opt.breakdown.silica:2008}.
 LAL is used to prepare colloidal solutions of nanoparticles (NP).
 These processes have many technological applications, see detailed reviews \cite{Stephan:2017.review,review.XIAO:2017}.
 Feeling about the today's level of achievements and unsolved problems may be obtained reading approximately hundred one-page abstracts from the ANGEL 2018 conference \cite{ANGEL}.


 In spite of numerous applications and many peoples involved in working in this direction, a lot of problems remain poorly understood.
 This is because the ultrafast early stages still are unobservable experimentally.
 Only rather late stages starting from the microsecond $(\mu$s) time scale are carefully documented \cite{Amans:Rayl-Plst:cav.bubb:APL:2016}.
 These late stages relates to formation and development of a cavitation bubble.
 While pulse durations from 0.1 ps to a few ns are from 7 to 3 orders of magnitude shorter.


 Recently computer simulations shed light onto the beginning of the "hidden era" lasting up to $t<1$ $\mu$s before the bubble begins to form
  \cite{Povarnitsyn-Itina:LAL:2013,povarnitsyn:ITINA:LAL:2014,LZ-bulk-LAL:2017,LZ+Stephan:2018.LAL,INA.arxiv:2018.LAL,INA.jetp:2018.LAL}.
 Ultrashort pulses with duration $\tau_L\sim 0.1$ ps were considered, and the processes have been followed up
   to few nanoseconds in \cite{Povarnitsyn-Itina:LAL:2013,povarnitsyn:ITINA:LAL:2014,LZ-bulk-LAL:2017,LZ+Stephan:2018.LAL}
     and up to 200 ns in \cite{INA.arxiv:2018.LAL,INA.jetp:2018.LAL}.
 In all these papers the absorbed energies $F_{abs}$ were comparable and of the order of $\sim 0.5$ J/cm$\!^2.$
 In the present paper we analyze new case with the three order of magnitudes longer pulse.

\begin{figure}       
\centering \includegraphics[width=0.75\columnwidth]{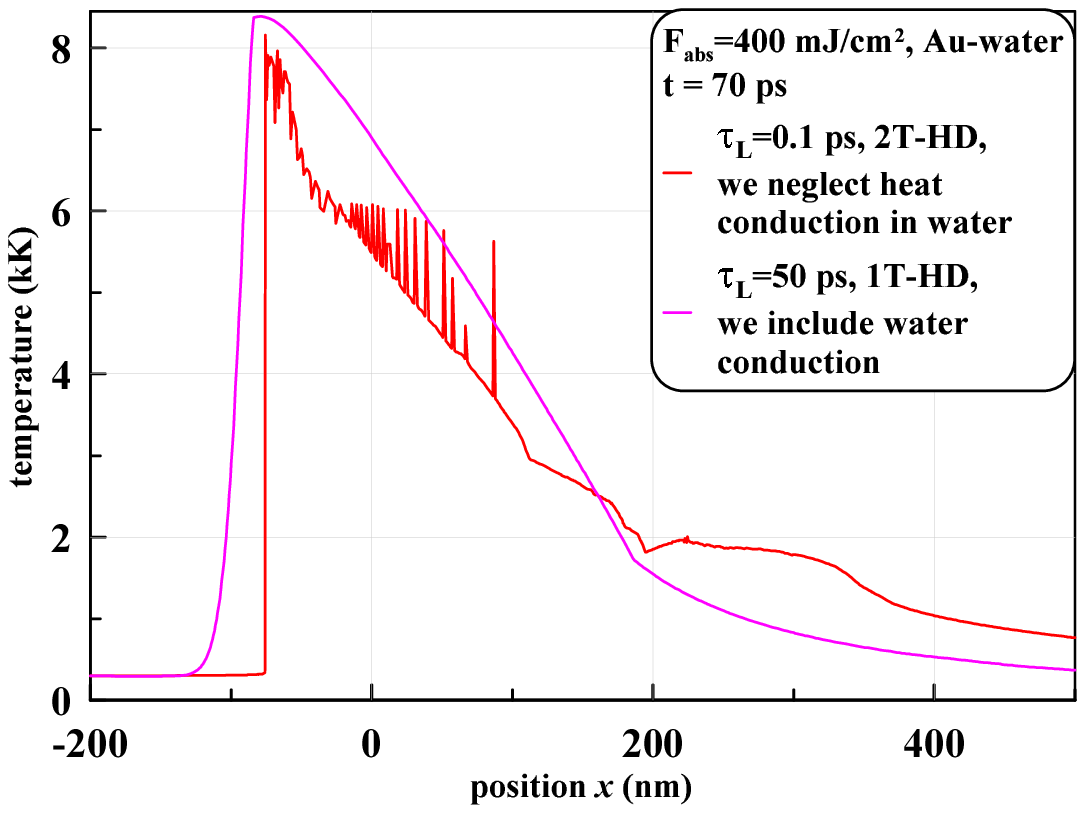}
\caption{\label{fig:01} Temperature distributions for two cases very different in pulse durations.
For ultrashort pulse the two-temperature (2T) hydrodynamics (2T-HD) code is used \cite{INA.arxiv:2018.LAL,INA.jetp:2018.LAL}
 and the profile of ion temperature $T_i(x,t=70\,\un{ps})$ is shown.
In case of subnanosecond pulse $\tau_L=50$ ps the 2T effects are of minor significance.
Therefore the one-temperature (1T) hydrodynamics (1T-HD) code is employed; the profile $T(x,t=70\,\un{ps})$ is presented;
 in the 1T approximation we have $T=T_i=T_e.$
Thermal conductivity of water is included in the 1T-HD code.
Temperature and heat flux are continue at the contact boundary (CB) between gold (Au) and water (wt).
Thermal diffusivity $\chi_{wt}$ of water is more than two orders of magnitude less than diffusivity of molten Au $\chi_{Au}.$
Therefore only a thin layer of water near CB is heated to the instant shown.
}
\end{figure}


 At first glance, it seems surprising that dynamics at the stage beginning after instant $t=50$ ps (when the 50 ps pulse is ended)
  is similar in the cases with so different durations $\tau_L$ (0.1 ps versus 50 ps) if energies $F_{abs}$ are the same (in more details this question is discussed below).
 This is what we find comparing the cases with $\tau_L=0.1$ ps and $\tau_L=50$ ps, see Fig. \ref{fig:01}.
 Thermal diffusivity $\chi_{Au}$ of gold is large at the 2T stage \cite{INA.jetp:2018.LAL} relative to its 1T values in solid and liquid states.
 But duration of the ultrashort pulse and duration of a 2T stage are shorter than duration 50 ps of the subnanosecond pulse.
 Therefore thicknesses of the heat affected zones are approximately the same in both cases, see Fig. \ref{fig:01}.


 Thicknesses of the liquid layers are approximately the same in Fig. \ref{fig:01}.
 The thin layer separating liquid and solid has approximately the position $x\approx 200$ nm in both cases at the instant shown.
 Our model of heat conduction includes the change of thermal conductivity caused by melting.
 Therefore the gradient $\nabla T$ is 2-3 times smaller in the solid.
 For the same reason the gradient $\nabla T$ has a sharp kink at the Au-water contact boundary (CB), see Fig. \ref{fig:01}.


 As was said, the thermal distributions are similar in the cases with the durations $\tau_L$ 0.1 ps and 50 ps shown in Fig. \ref{fig:01}.
 But there are differences also.
 Before saying about them let's describe a global structure of the 1D flow.
 In the both cases the foam region and atmosphere are formed
   \cite{Povarnitsyn-Itina:LAL:2013,povarnitsyn:ITINA:LAL:2014,LZ-bulk-LAL:2017,LZ+Stephan:2018.LAL,INA.arxiv:2018.LAL,INA.jetp:2018.LAL}.
 The atmosphere is a layer contacting with water and being in quasi-hydrostatic equilibrium with CB as it is described in \cite{INA.arxiv:2018.LAL,INA.jetp:2018.LAL}.
 The atmosphere separates foam and water.
 In the case of a rather long subnanosecond pulse the atmosphere forms when CB begins to decelerate.
 As it will be seen below, the CB accelerates during the rise of the pulse $I\propto \exp(-t^2/\tau_L^2), \tau_L=50$ ps and some time after;
   we begin our simulations at the negative time instant equal to $-3\tau_L.$


 In the ultrashort case shown in Fig. \ref{fig:01} the nucleation process and formation of foam begins earlier than in the subnanosecond case.
 This is the difference between the cases in Fig. \ref{fig:01}.
 In the ultrashort case nucleation begins at the stage $t\approx 20$ ps while in the subnanosecond case the starting stage of nucleation is $t\approx 80$ ps.




 Nevertheless the amount of the ejected gold mass $m_{ej}\approx (1.5-3)\cdot 10^{-4}$ g/cm$\!^2$ is approximately the same in the both cases.
 The ejected mass forms a layer separated by a gap from the rest of a bulk gold target up to the stages $\sim 10^2$ ns.
 The layer contacts with water.
 Velocity of the layer drops down to very small values $\sim 1-10$ m/s at the stage $\sim 10^2$ ns \cite{INA.arxiv:2018.LAL,INA.jetp:2018.LAL}.
 Fate of the layer is unknown at the late stages $\geq \mu$s.
  The gap is filled by saturation vapor of gold and has pressures $\sim 10^2-10^3$ bars until gold has temperatures above $\approx 5$ kK \cite{INA.arxiv:2018.LAL,INA.jetp:2018.LAL}.
 While pressure in liquid and later in gaseous water is $\sim 10^2-10^3$ bars at the stage of few microseconds.
 Thus the gold layer may moves back under water pressure and closes the gap if the layer is cooled down below few kK.
 The layer may stick back with the bulk gold if gold in the layer remains liquid during the shrinkage of the gap.
  Velocity change $\Delta v$ of the layer under pressure difference $\Delta p$ acting onto the layer from its internal (vapor of Au in the gap) and external (CB, water) boundaries is
    large $\Delta v = \Delta p \cdot t_{\mu s}/m_{ej}\sim 3$ km/s for $\Delta p=10^2$ bars,
     $m_{ej}= 3\cdot 10^{-4}$ g/cm$\!^2,$ and $t_{\mu s}\sim 1$ $\mu$s.
 Such velocities easily may close the gap during $\mu$s because the gap thickness is $\sim \mu$m only.
 This problem obviously needs separate consideration.


 Below we will limit ourself with comparative analysis of the early $(\sim $10s ps) and middle $(\sim 1$ ns) stages
   of the ultrashort Au-water and subnanosecond Au-water and Au-glass cases.
 There are two main conclusions following from this analysis.

 First. All these three cases are similar in the sense of (i) energy distribution in the surface layer, (ii) the value of ejected mass,
   and (iii) the trajectory of the CB $\zeta (x=x_{CB}(t), t),$ where the $\zeta$ stands for kinematic $(x,v,g$ shift, velocity, acceleration)
     or for thermodynamics $(\rho, p, T)$ quantities.

 Second. The states near the CB during the middle stage $\sim 1$ ns are hot and compressed, thus both the gold and, of course, the water (with its low critical parameters)
  are in overcritical states.
 Therefore Au-water surface tension coefficient is zero and diffusion is strong.
 Rather thick diffusively mixed Au-water layer forms around the CB thanks to that.
 Later in time the pressure drops due to expansion of the pressurized layer near the CB.
 Condensation into clusters and liquid droplets begins as pressure decreases below the critical values for gold.


 From the other hand, strong diffusion suppresses development of the Rayleigh-Taylor instability (RTI), see Section 7 below.
 There is a limited in time interval when the RTI develops \cite{INA.arxiv:2018.LAL,INA.jetp:2018.LAL}, see Section 3 and Fig. \ref{fig:11}.
 Thus if the development is delayed during this temporal interval then contribution of RTI remains insignificant.
 Significant contribution from RTI into the nanoparticles (NP) production was observed for ultrashort actions and smaller (than in this paper) absorbed energies in
    \cite{LZ-bulk-LAL:2017,LZ+Stephan:2018.LAL,INA.arxiv:2018.LAL,INA.jetp:2018.LAL}.



 The paper is organized as follows.
 First in Section 2 we consider separation of large distances covered by shocks in gold and water from small distance passed by contact boundary (CB).
 Velocity of a CB decays in time quickly while shocks propagate at velocities above speed of sound.
 In Section 3 the acceleration stage is considered.
 This stage differs the case of ultrashort pulse from the case of subnanosecond pulse.
 Later in time (that is after the acceleration stage) these cases behave similarly.
 In Section 4 duration of the stage when parameters of a metal at the CB are in overcritical states (overcritical stage) is analyzed.
 It is found that this duration is a function of absorbed energy.

 We use two-temperature (2T) and one-temperature (1T) hydrodynamic codes to obtain results listed in the Sections 2-4.
 Embedded atom method (EAM) potentials of interatomic interactions and molecular dynamics (MD) simulations are applied
   to reveal another part of valuable information in the next Sections.
 Saturation pressure, curves of equilibrium between phase states, surface tension coefficient, and description of diffusion on the base of the EAM/MD approach
   is presented in Section 5.
 Condensation of gold and formation of nanoparticles is described in Section 6.
 Section 7 demonstrates results of large scale MD simulation of ablation of gold in water.

\begin{figure}       
   \centering   \includegraphics[width=0.75\columnwidth]{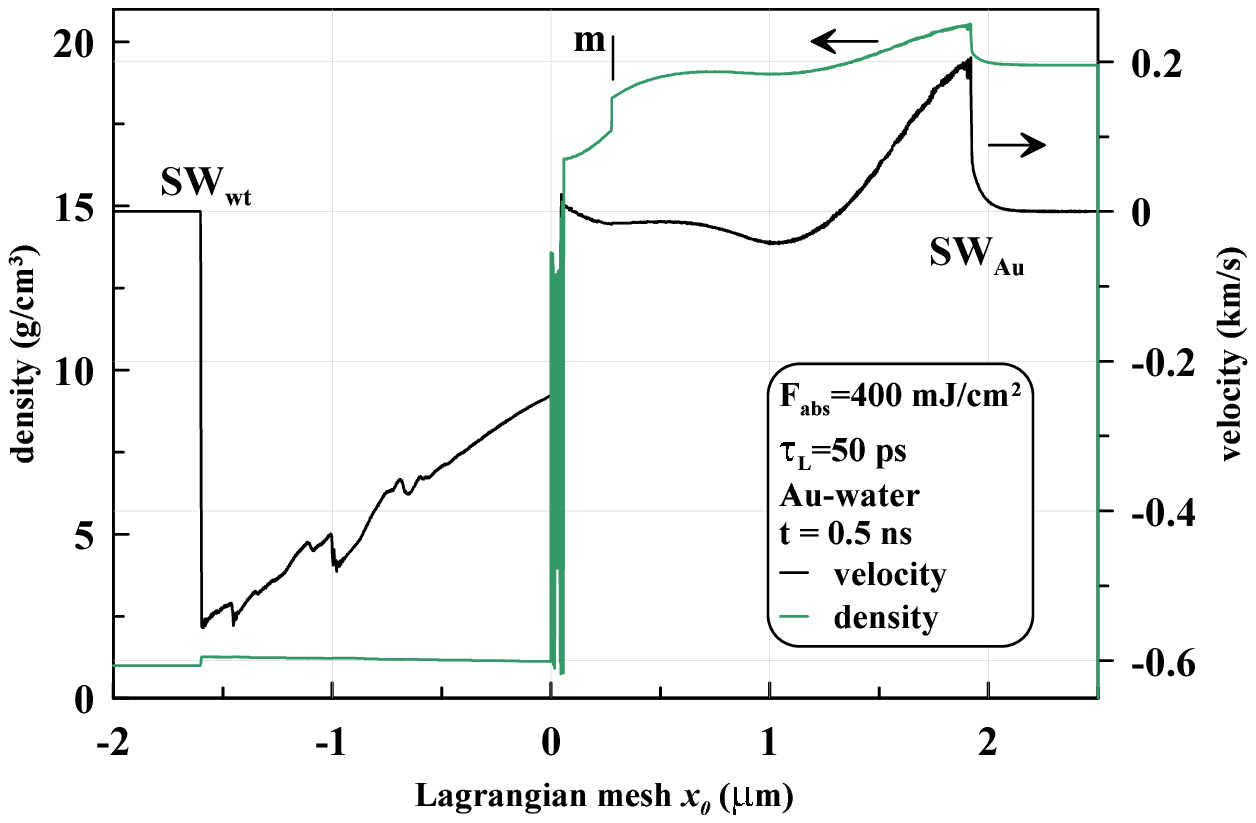}
\caption{\label{fig:02} Density and velocity profiles are shown.
The point "m" marks current position of a boundary between solid and liquid gold.
We see that at the stage after $\sim 1$ ns the near contact layer is much more thinner than the region between the two SWs.
 The profiles are plotted as functions of initial positions $x_0$ of material particles of gold and water.
 Thus reader can easily estimates masses.
 E.g., current mass of shock compressed water is $x_0|_{SW-wt}*\rho_0|_{wt},$ where $x_0|_{SW-wt}(t)$ is position of the SW in water, $\rho_0|_{wt}$ is initial density of water.
 Lagrangian positions are motionless; $x_0=0$ is the CB.
  }  \end{figure}

\begin{figure}       
   \centering   \includegraphics[width=0.75\columnwidth]{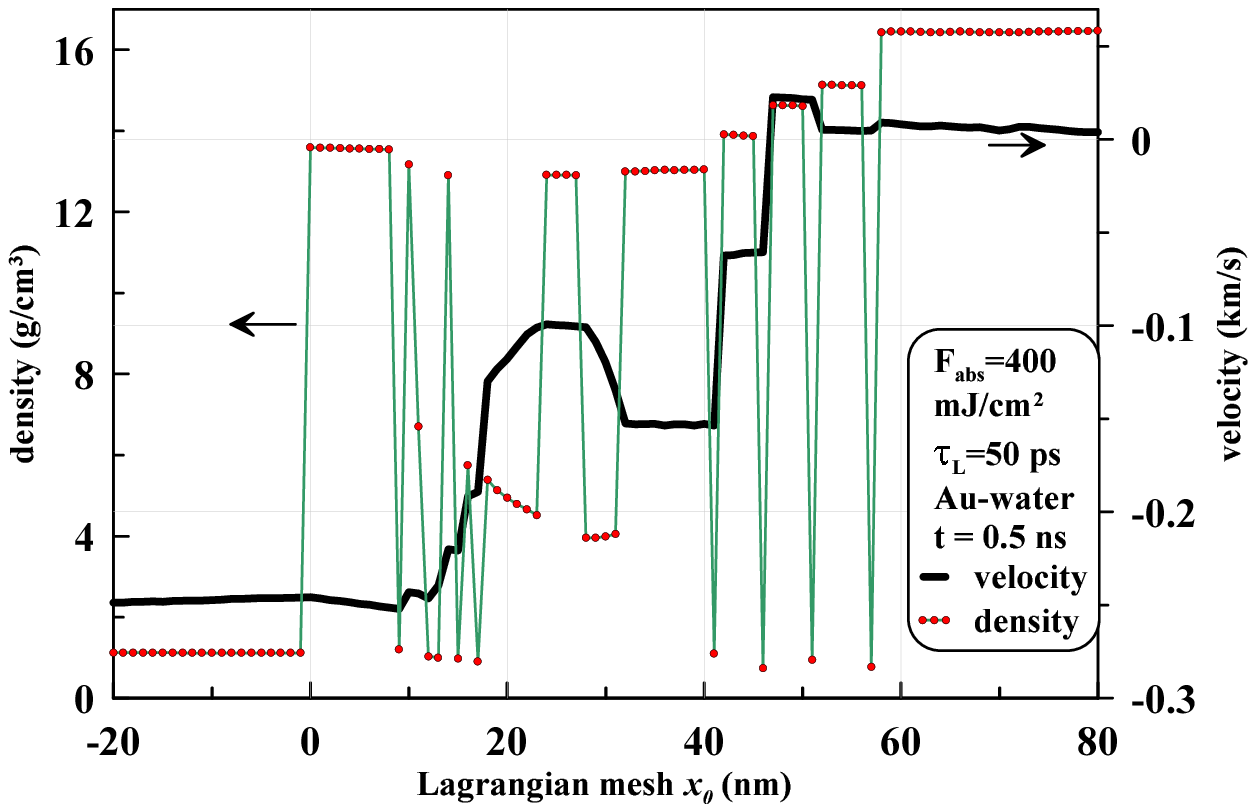}
\caption{\label{fig:03} Internal structure of the foamy layer near the CB from previous Fig. \ref{fig:02}.
We see how the velocity "jump" from Fig. \ref{fig:02} is accumulated along the stretching foamy layer.
Inertia of gold supports this stretching at the temporal stage shown.
Small red circles mark positions of the steps of the Lagrangian mesh.
The mesh should cover large distances passed by SWs.
Nevertheless it satisfactory describes foaming - there are few mesh points per every liquid peace of the foam.
  }  \end{figure}

\section{Global and internal structures}




 Global structure of flow is clear from Fig. \ref{fig:02}.
 Before action of a pulse we have two motionless half-spaces in contact.
 The left one $x_0<0$ is filled with water, while the right one $x_0>0$ corresponds to the bulk gold target;
   here $x_0$ is Lagrangian coordinate; it corresponds to initial (before laser action) positions of material particles.
 Laser beam penetrates through transparent water and heats absorbing gold in its skin-layer.
 Heating and pressure rise cause motion.
  After finishing of a pulse the spatial expansion gradually decreases pressure.
 Two shock waves (SW) in Fig. \ref{fig:02} run into water and into gold target.
 The pulse is rather long $\tau_L=50$ ps.
 Therefore it takes some time to form the shocks.
 The SWs form thanks to nonlinear effects after focusing of characteristics and wave breaking.

\begin{figure}       
   \centering   \includegraphics[width=0.7\columnwidth]{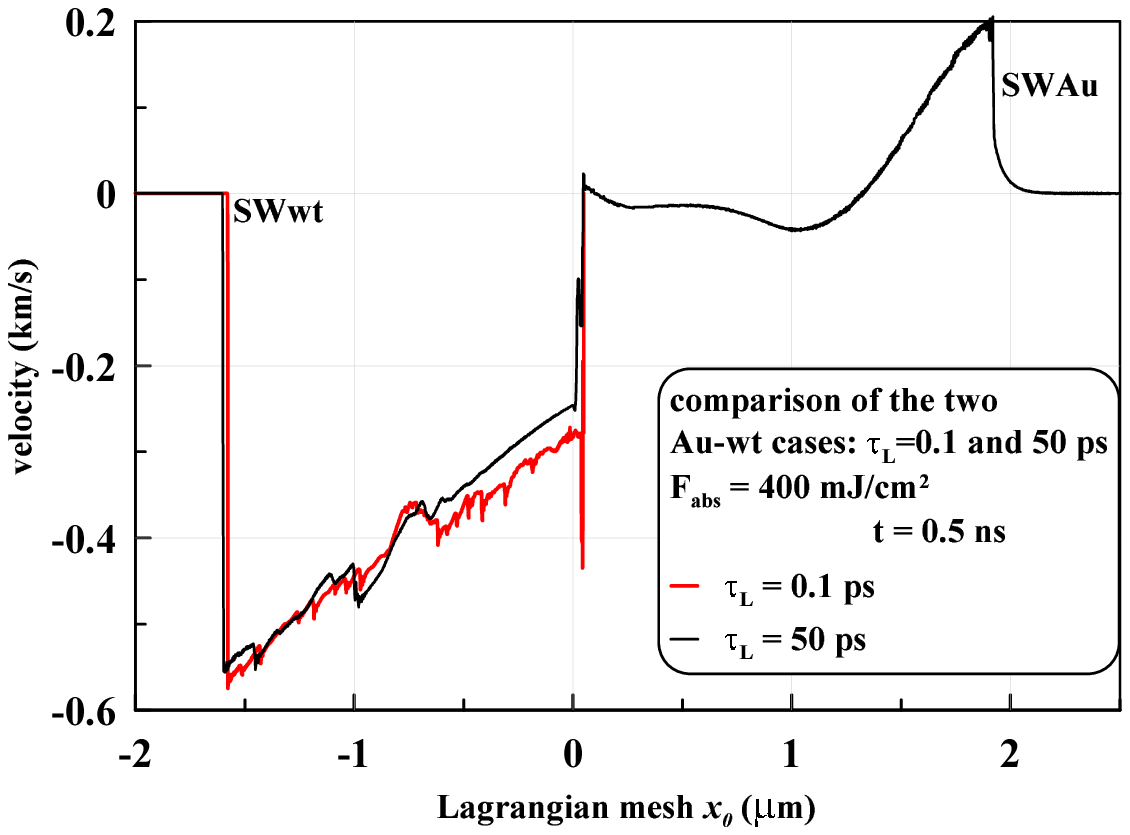}
\caption{\label{fig:04} Comparison of the velocity profiles for two very different durations of laser action.
The velocity profile for the simulation with $\tau_L=50$ ps is taken from Fig. \ref{fig:02}.
  }  \end{figure}


 Velocity "jump" 0.23 km/s in Fig. \ref{fig:02} covers the layer where the atmosphere and foam locate.
 The "jump" characterizes expansion of the CB relative to the bottom of the foam at the instant 0.5 ns.
 This velocity is small (just few \%) in comparison with velocity $c_s|_{Au}+c_s|_{wt}=4.5$ km/s giving the lowest limit of velocity
   of increase of a distance between the SWs in Fig. \ref{fig:02}.
 Expansion proceeds as a result of easy stretchability of a vapor component in foam.
 Structure of the thin layer near the CB (the "jump" in Fig. \ref{fig:02}) is revealed in Fig. \ref{fig:03}.

\begin{figure}       
   \centering   \includegraphics[width=0.7\columnwidth]{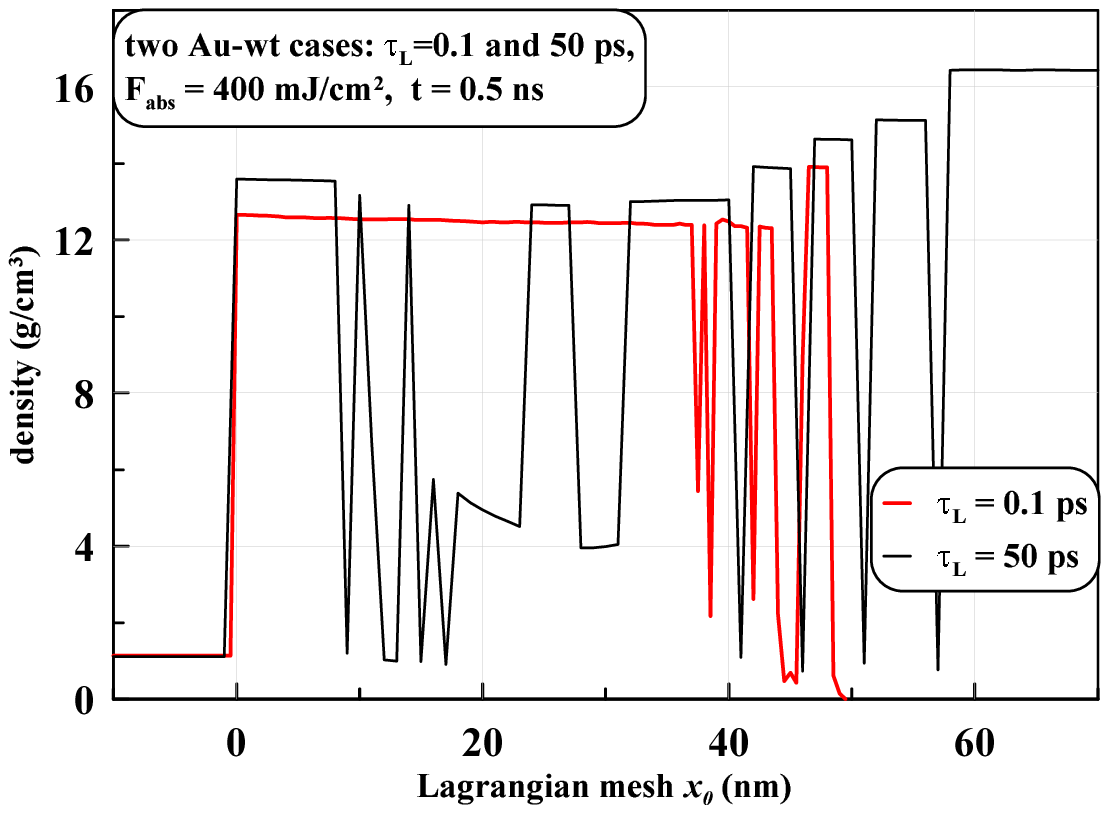}
\caption{\label{fig:05} Density structure in foamy region for the two cases compared in Fig. \ref{fig:04}.
  }  \end{figure}


 Comparison of the Au-water simulations for ultrashort pulse $\tau_L=100$ fs \cite{INA.arxiv:2018.LAL,INA.jetp:2018.LAL}
  and subnanosecond pulse with $\tau_L=50$ ps is demonstrated in Fig. \ref{fig:04}.
 Absorbed energies are the same.
 We cut out the slowly moving part of bulk gold in the ultrashort case \cite{INA.arxiv:2018.LAL,INA.jetp:2018.LAL}.
 Only really significant for mixing with water and nanoparticles production part of a gold target was kept in the simulation of ultrashort action.
 Comparing two simulations in Fig. \ref{fig:04} we conclude that:

 First, at the nanosecond time scale and later the duration varied 500 times from 0.1 to 50 ps is insignificant.

 Second, the passive part of the gold target kept in one simulation (pay attention to the SW running into bulk)
 and removed in the another one is, indeed, insignificant for dynamics of foam and water.

\begin{figure}       
   \centering   \includegraphics[width=0.7\columnwidth]{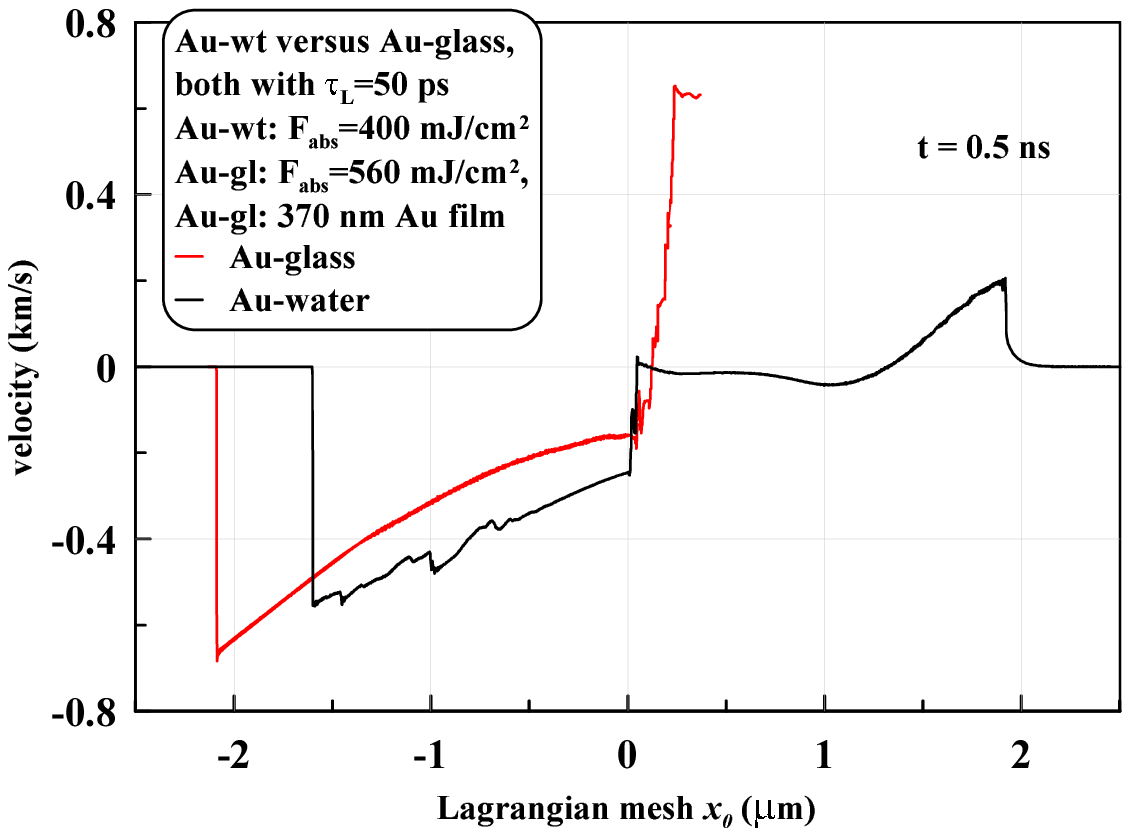}
\caption{\label{fig:06} Velocity profiles for the instant 0.5 ns for the Au-wt and Au-glass cases of the same duration 50 ps.
  }  \end{figure}

\begin{figure}       
   \centering   \includegraphics[width=0.7\columnwidth]{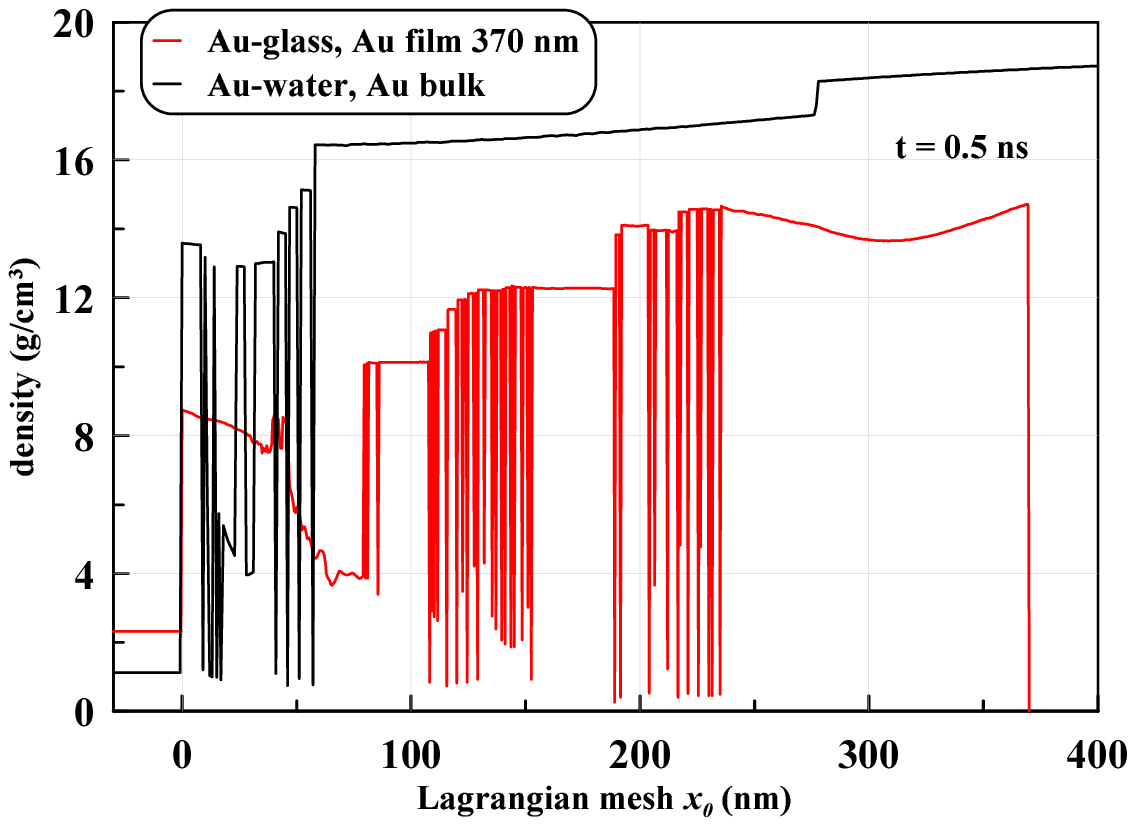}
\caption{\label{fig:07} Density profiles near the CB for the same two cases as in Fig. \ref{fig:06}.
  }  \end{figure}


 But internal structures of density distributions inside the foamy layer differ, see Fig. \ref{fig:05}.
 This is because of random origin of nucleations and separation into vapor and liquid phases
  and (may be) due to different acceleration/deceleration behavior at the stage of the order of a few tens of ps.



 Expansions in water (liquid dielectrics) and in glass (initially solid dielectrics) qualitatively are similar
    up to the stage when in the case of water the bubble begins to form.
 Corresponding situations are shown in Fig. \ref{fig:06}.
 In Figures \ref{fig:06} and \ref{fig:07} the cases with a bulk gold target and a thick film target (initially 370 nm thick) are compared.
 We call a film thick if its thickness is above thickness of a heat affected zone (HAZ) \cite{Inogamov2015nanoBump,PhysRevApplied:2017}.
 Thickness 370 nm is larger than thickness of HAZ, see Fig. \ref{fig:01}.
 The case with a film 370 nm and a glass substrate is considered in connection with experiments \cite{Li:17,Li2017}.
 Discussion in connection with these experiments needs separate description.
 Figures \ref{fig:06} and \ref{fig:07} demonstrate dramatic difference between the cases with a bulk and a film targets if we consider motion of gold;
   dynamics of dielectrics are similar as it follows from Fig. \ref{fig:06}.
 Initially the system from dielectrics and a target was motionless, thus its momentum is zero.
 Laser beam introduces energy $F_{abs}$ into the system, but not momentum because momentum of a photon cloud is negligibly small.
 Therefore total momentum of a system is zero after laser impact.


 The momentum directed into the side of dielectrics produces motions in dielectrics which are more or less similar
  in both cases, see Fig. \ref{fig:06}.
 But the motions in the gold targets triggered by the right side momentum are very different
   depending on thickness of a film.
 In a bulk target this momentum is carried into the bulk side and out from the near CB layer
   by approximately triangular shock wave SW$_{Au}$ in Figures \ref{fig:02}, \ref{fig:04}, and \ref{fig:06}.
 In the case of a film this carryover is impossible - thickness is limited.
 Thus all amount of the powerful push to the right remains in a film.
 This causes decay of a film into a sequence of quasi-spallative-plates separated by foamy layers in case of a film, see Fig. \ref{fig:07}.
 Velocity of the rear-side plate is large, see Fig. \ref{fig:06}.


\begin{figure}       
   \centering   \includegraphics[width=0.75\columnwidth]{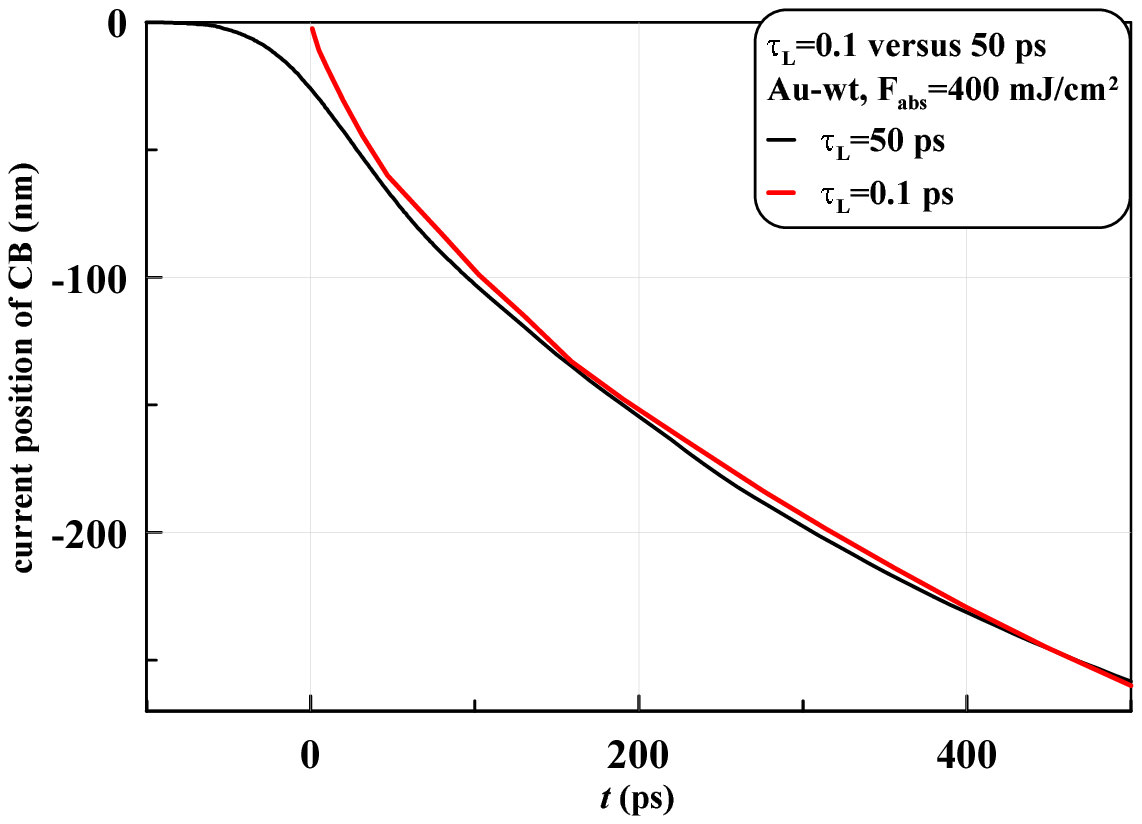}
\caption{\label{fig:08} In this picture we compare the ultrashort and subnanosecond actions onto the Au-water system.
Thus this picture is similar to Fig. \ref{fig:04}.
We use the same colors for these two cases as in Fig. \ref{fig:04}.
Let's emphasize remarkable similarity of these cases as it was already mentioned in Fig. \ref{fig:04}.
The trajectories are very similar if we exclude the speed up interval $t<0,|t|\sim \tau_L=50$ ps
 for subnanosecond drive.
  }  \end{figure}

\section{Active dynamics: acceleration, deceleration, and comparisons}




 Dynamics of expansion of CB in the case of Au-wt pair is shown in Fig. \ref{fig:08}.
 The cases with ultrashort and subnanosecond pulses are compared.
 Absorbed fluences are the same 400 mJ/cm$\!^2.$
 In both cases intensity is given by the Gaussian law $I\propto \exp(-t^2/\tau_L^2).$
 But durations $\tau_L$ differ 500 times.
 Thus if we impose these Gaussian laws onto the plane in Fig. \ref{fig:08}
  then the subnanosecond pulse will have visible left and right wings.
 And the rise of its left wing is proportional to the growing shift of CB in the subnanosecond case.
 While the Gaussian for the ultrashort pulse looks like delta-function at the temporal scale used in Fig. \ref{fig:08}.


 The speed up interval is very short for the pulse with $\tau_L=0.1$ ps.
 We don't see it because the time scale in Fig. \ref{fig:08} is too large.
 At this scale, it seems that in the ultrashort case the CB starts to move immediately at the instant $t=0$ with large velocity.
 Later in time $t>0$ it only decelerates thanks to resistance of water to expansion of gold.
 But in the subnanosecond case the definite acceleration stage exists.
 Of course, it is related to the increase in time of absorbed intensity at the left wing of Gaussian pump.

\begin{figure}       
   \centering   \includegraphics[width=0.75\columnwidth]{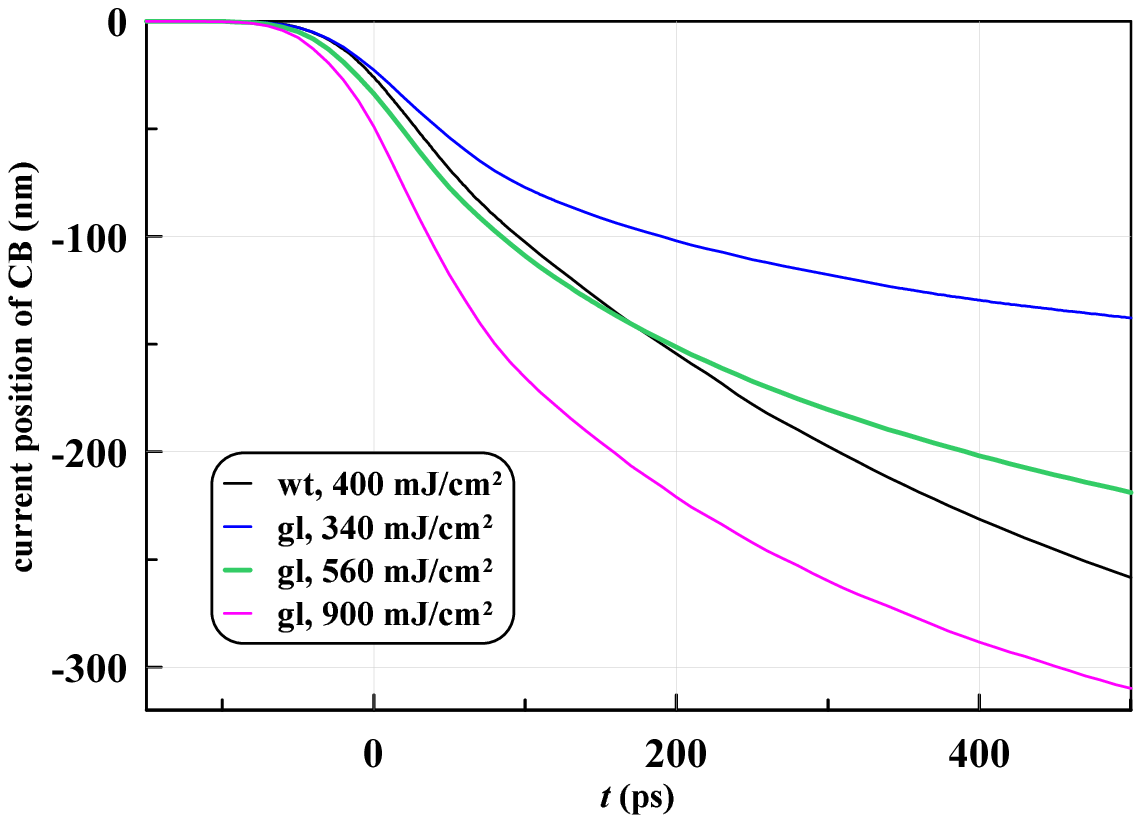}
\caption{\label{fig:09} Trajectories of CB in four cases with a nanosecond drive.
Abbreviations "wt" and "gl" relate to the Au-water and Au-glass systems.
Digits in mJ/cm$\!^2$ give absorbed energies.
There is a gold film 370 nm thick in the Au-glass cases.
We use bulk gold target in the Au-wt case.
  }  \end{figure}


 Influence of absorbed energy, resistive ability of dielectrics, and thickness of a gold target on the trajectory of the contact boundary (CB) $x_{CB}(t)$
   is illustrated in Fig. \ref{fig:09}.
 Obviously, the shift increases if we increase absorbed energy fixing two other parameters: density of dielectrics
   and thickness of a gold layer.
 If we increase density of surrounding medium then deceleration is faster;
   initial density of the pyrex silica used in our Au-glass simulations is 2.2 g/cm$\!^3.$
 Compare the cases wt/400 mJ/cm$\!^2$ and gl/560 mJ/cm$\!^2$ in Fig. \ref{fig:09}.
 This is also obvious.


 Less obvious are consequences of variation of thickness of a layer of Au.
 It seems, at first sight, that a thick target will decelerate more slower.
 But there is foam formation and appearance of a contact layer of gold (atmosphere) \cite{INA.arxiv:2018.LAL,INA.jetp:2018.LAL}.
 Speed of sound in foam is low.
 The first portions (atmosphere) of fragmented Au are decelerated by surrounding water or glass,
  while the second portions move faster thus adding their mass to the first portions located at the CB.
 This is said to underline that (i) there is a decelerated layer (atmosphere) and
  (ii) its mass $m_{ej}$ may change with time.

\begin{figure}       
   \centering   \includegraphics[width=0.75\columnwidth]{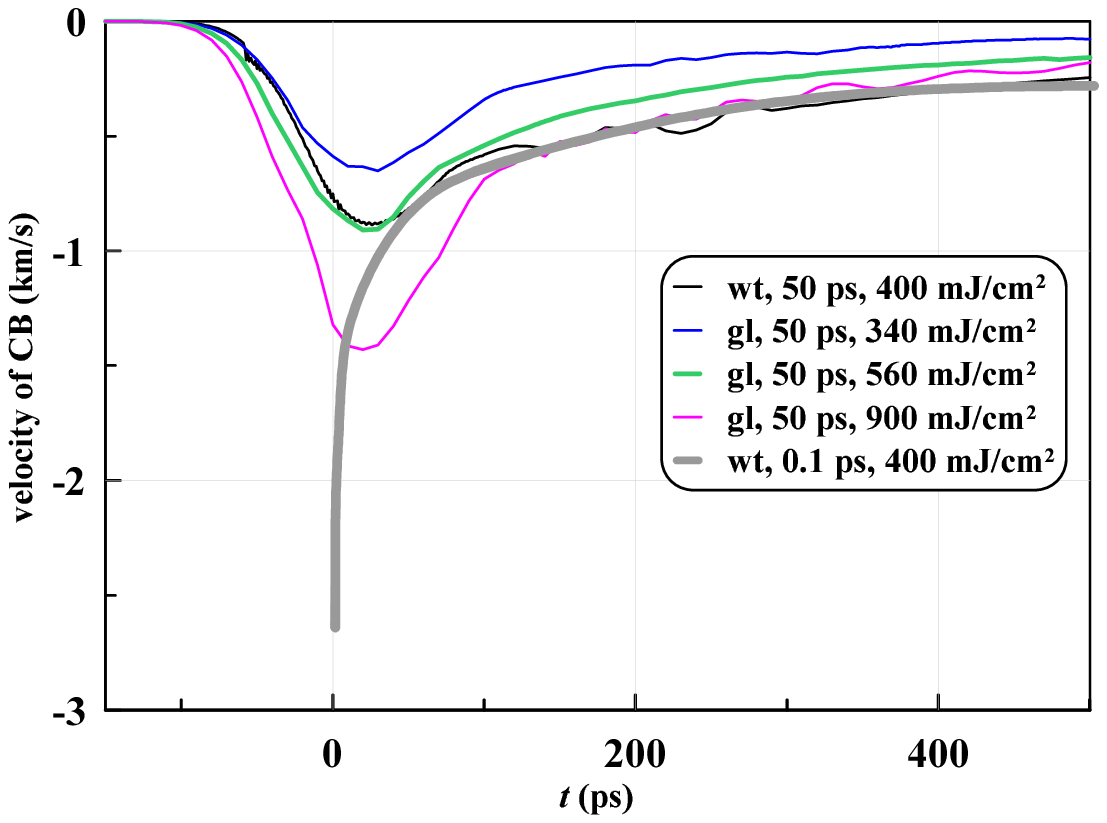}
\caption{\label{fig:10} Velocities of contact boundary (CB) $v_{CB}(t).$
 See designations in Fig. \ref{fig:09}. The digits in ps are the e-folding durations of pulses.
 The picture emphasizes qualitative distinctions between ultrashort and subnanosecond pulses at the early stage.
 The delay of beginning of deceleration significantly influences development of Rayleigh-Taylor instability in direction of suppression.
  }  \end{figure}


 Analysis of simulations show that asymptotically velocity $v$ and pressure $p$ at the CB
   are acoustically connected $p=Z v,$ where $Z=\rho c_s$ is acoustic impedance in surrounding medium.
 Writing $m_{ej} \ddot x=-Z \dot x$ we find that $\dot x \propto \exp(-t/t_{dec}),$
  \begin{equation}\label{eq:t-decelerate}
  t_{dec}=m_{ej}/Z\approx (\rho_{Au}/\rho)h_{ej}/c_s,
  \end{equation}
   where $x$ is CB position, $\rho$ and $c_s$ are density and speed of sound of surrounding matter.
 The estimate (\ref{eq:t-decelerate}) gives for the deceleration time constant a value $t_{dec}=0.7$ ns for density ratio $\rho_{Au}/\rho\approx 10,$
  height of atmosphere $h_{ej}\approx 100$ nm, and speed of sound in water $c_s=1.5$ km/s.


 Another important conclusion from Figures \ref{fig:08}, \ref{fig:09}, and the estimate (\ref{eq:t-decelerate}) for the deceleration time constant $t_{dec}$
   is that in all cases the shift $x_{CB}$ is very limited - it is just of the order of micron.


 The difference between ultrashort and subnanosecond laser drives is distinctly clear from Fig. \ref{fig:10}.
 Whereas there is a relatively prolonged acceleration interval in the subnanosecond case, in the ultrashort case the velocity trajectory starts from high value
  and after that only decelerates with time.
 Let's mention that acceleration continues after the instant $t=0$ when the maximum of a laser pump is achieved.
 Thus intensity begins to decrease while velocity still increases.
 Increase of velocity continues approximately $t_{max-vel}=20-25$ ps after maximum of a pump in $t=0.$

\begin{figure}       
   \centering   \includegraphics[width=0.75\columnwidth]{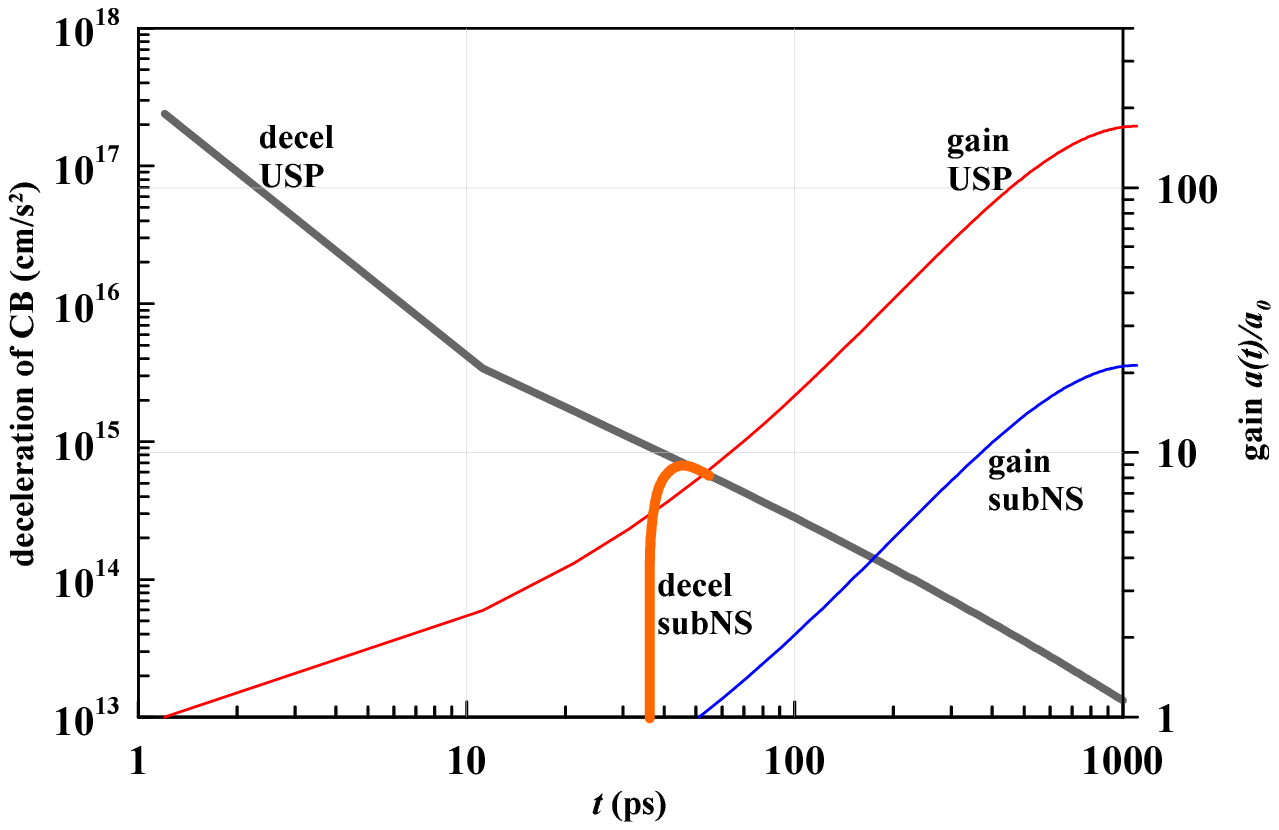}
\caption{\label{fig:11} Deceleration of contact boundary (CB) for the ultrashort pulse (USP) is marked as "decel USP"
 and as "decel subNS" for the subnanosecond pulses.
 In the subnanosecond case the acceleration is changed by deceleration in the instant $t_{max-vel},$ see Fig. \ref{fig:10}.
 After the time interval $\sim 10$ ps long after the instant $t_{max-vel}$ the dependence $v_{CB}(t)$ asymptotically tends to the dependence for the ultrashort pulse.
 Thus the deceleration for the subnanosecond pulse asymptotically approximately tends to the deceleration in case of the USP.
 This is shown by the curve "decel subNS" starting from zero and going approximately to the $\ddot x_{CB}$ for USP.
 This difference in deceleration trajectory causes delay in the Rayleigh-Taylor gain,
  compare the curves "gain USP" and "gain subNS" shown, these curves relate to the right vertical axis, see text for explanations.
  }  \end{figure}


 Velocity drop in the ultrashort case shown in Fig. \ref{fig:10} is well described by a function $v=4.05\exp(-0.63 (t-1.1)^{0.23})$ km/s where time $t$ should be taken in ps.
 The subnanosecond velocity trajectories obey the approximately same law but only after few tens of ps after their maximum of velocity $t_{max-vel}.$
 Deceleration dependence on time is plotted in Fig. \ref{fig:11}.
 There is relatively slow acceleration in the subnanosecond case.
 The acceleration interval is lost for the Rayleigh-Taylor (RT) amplification of perturbations of CB.
 Thus development of the RT instability is significantly weaker in the subnanosecond case, compare the curves "gain USP" and "gain subNS" in Fig. \ref{fig:11}.
 The temporal interval of a few tens of ps is important for development of the RT disturbances
   because during this interval the deceleration is huge, see the curve "decel USP" in Fig. \ref{fig:11}.
 These decelerations overcome values $10^{14}$ cm/s$\!^2$ typical for surface of a neutron star.
 Atomic decelerations in interatomic interactions are of the order of $10^{18}$ cm/s$\!^2.$
 The curves "gain USP" and "gain subNS" in Fig. \ref{fig:11} corresponds to the wavelength 60 nm of a harmonic perturbation.
 They were calculated using linear RT theory including surface tension and viscosity according to the method described in \cite{INA.arxiv:2018.LAL,INA.jetp:2018.LAL}.

\begin{figure}       
   \centering   \includegraphics[width=0.7\columnwidth]{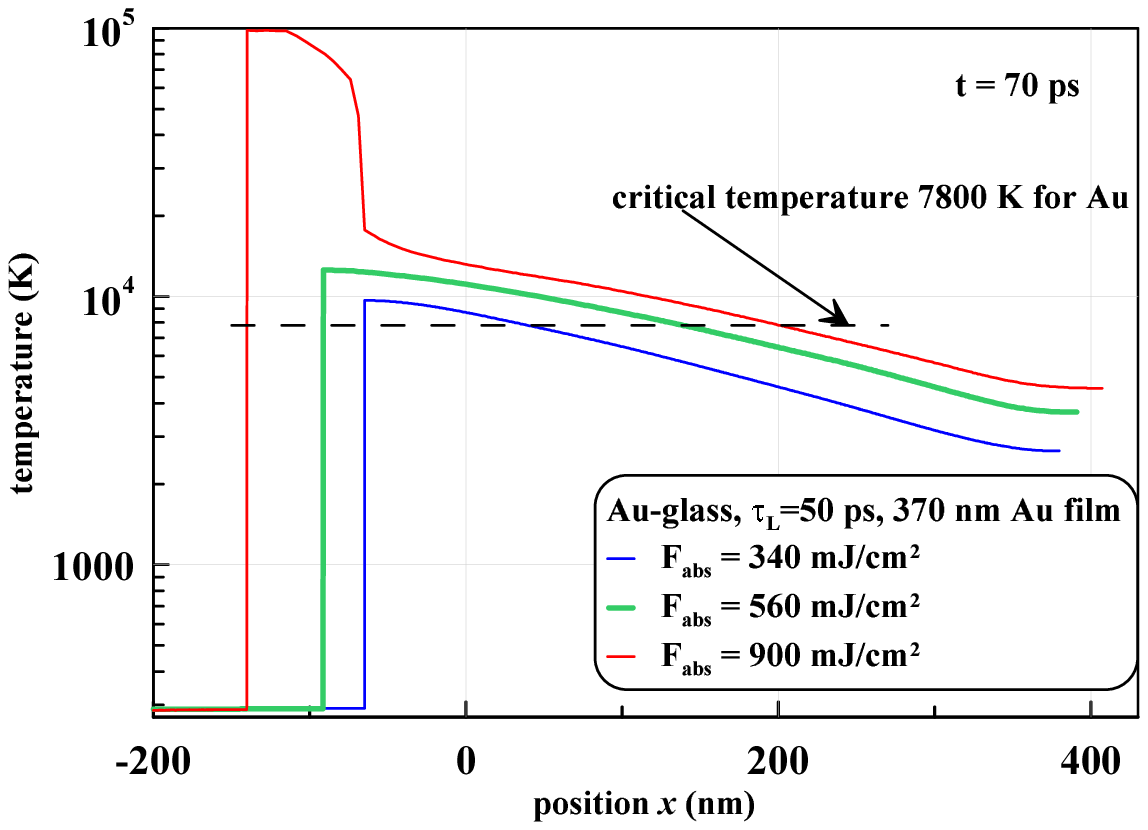}
\caption{\label{fig:12} Maximum heating after absorption of a subnanosecond pulse.
We see that for all three cases temperatures near CB are higher than critical temperature.
Comp. also with Fig. \ref{fig:01}.
  }  \end{figure}


\section{Strong heating, inertial geometrical confinement, overcritical states}


 Gold is strongly heated at the end stage of absorption of a subnanosecond pulse, see Figures \ref{fig:01} and \ref{fig:12}.
 Simulations are done using equation of state for gold, water and glass taken from \cite{Bushman:1993,Khishchenko2002,Lomonosov_2007,rusbank1,rusbank2}.
 We use heat conduction description for gold from \cite{INA.jetp:2018.LAL,Petrov.INA.KPM.jetp.lett:2013}.
 The description includes (a) calculated contribution from electron-electron collisions important at high temperatures and (b) decrease of conductivity during melting.
 Critical parameters of gold are 7.8 kK, 0.53 GPa, and 5.3 g/cm$\!^3$ \cite{Bushman:1993,Khishchenko2002,Lomonosov_2007,rusbank1,rusbank2}.

\begin{figure}       
   \centering   \includegraphics[width=0.7\columnwidth]{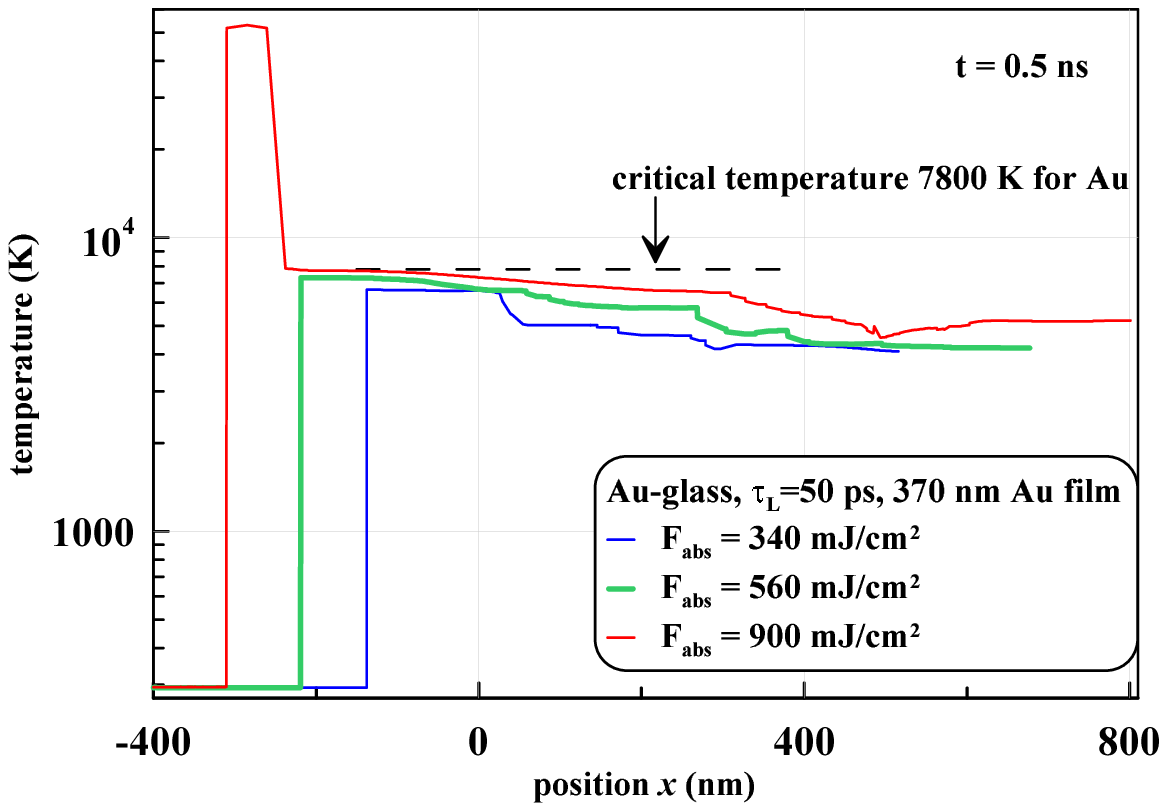}
\caption{\label{fig:13} Fig. Gradual cooling of gold (comp. with Fig. \ref{fig:12}) as a result of thermal conduction and work done for mechanical expansion of matter.
When the transition from the overcritical state to the two-phase states takes place(?) and how long gold near CB exists in the overcritical state(?) - these questions are important,
 because diffusive mixing is strongly enhanced in the overcritical states, thus more amount of gold is transferred into nanoparticles during cooling of a mixture
   since a diffusive layer is thicker.
 Temperature jumps at the profiles $T(x,t=500\,\un{ps})$ are connected with fragmentation and formation of a sequence of liquid layers, see Figures \ref{fig:03}, \ref{fig:05}, and \ref{fig:07}.
 Vapor layers in this sequence are less conductive than liquid.
  }  \end{figure}

\begin{figure}       
   \centering   \includegraphics[width=0.7\columnwidth]{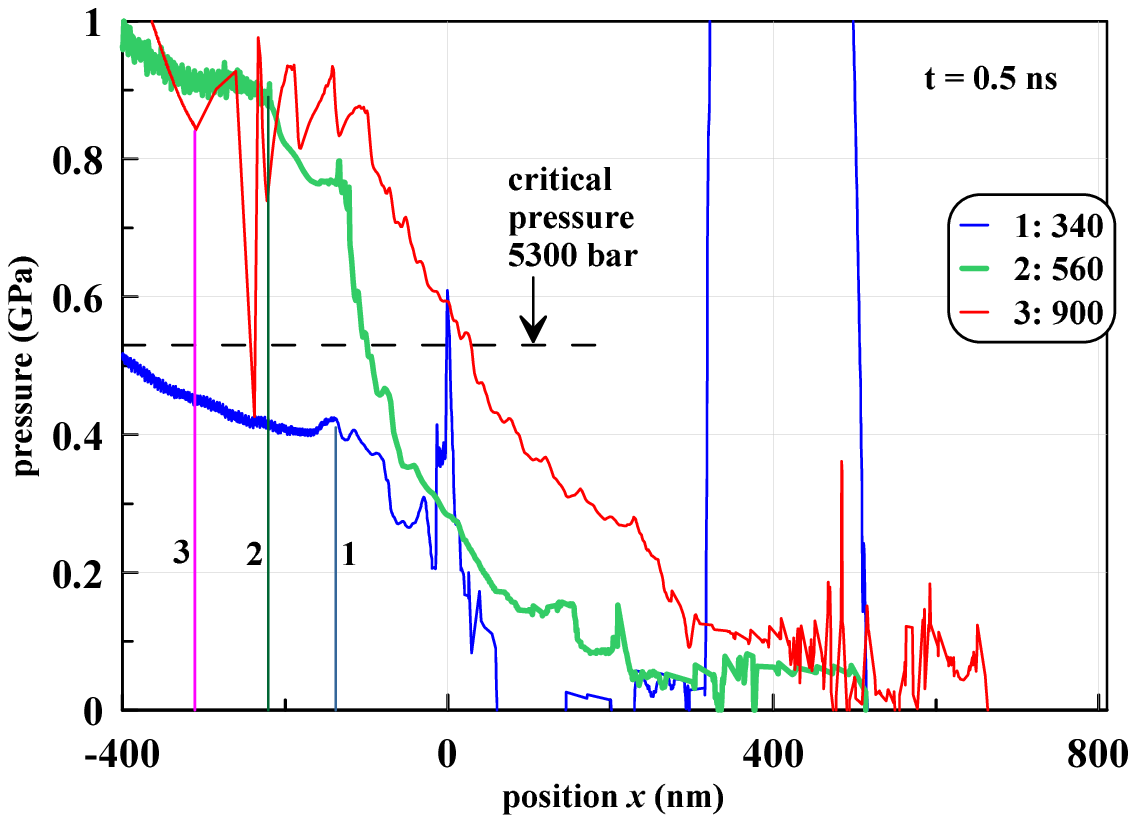}
\caption{\label{fig:14} Pressure profiles near the CB at the nanosecond time scale. We see that only in the case
with $F_{abs}=340$ mJ/cm$\!^2$ pressure drops below the critical value for Au. The vertical straight lines 1, 2, and 3
designate current positions of the CB for the three cases considered. Simulation parameters are the same as
in Figures \ref{fig:12} and \ref{fig:13}. The instant of time is the same as in Fig. \ref{fig:13}.
The CB separates dense (Au) and less dense (dielectrics) media in the cases 1 and 2;
correspondence between the digits 1, 2, and 3 and values of $F_{abs}$ is shown in the inset.
Therefore in these cases the $\nabla p$ has the kink at the CB. The layers of positive and negative $p$ (seen deeper to the right relative to the CB)
of rather significant amplitudes (few GPa) are caused by oscillations of the spallation plates
seen in Figures \ref{fig:03}, \ref{fig:05}, and \ref{fig:07}.
  }  \end{figure}

\begin{figure}       
   \centering   \includegraphics[width=0.7\columnwidth]{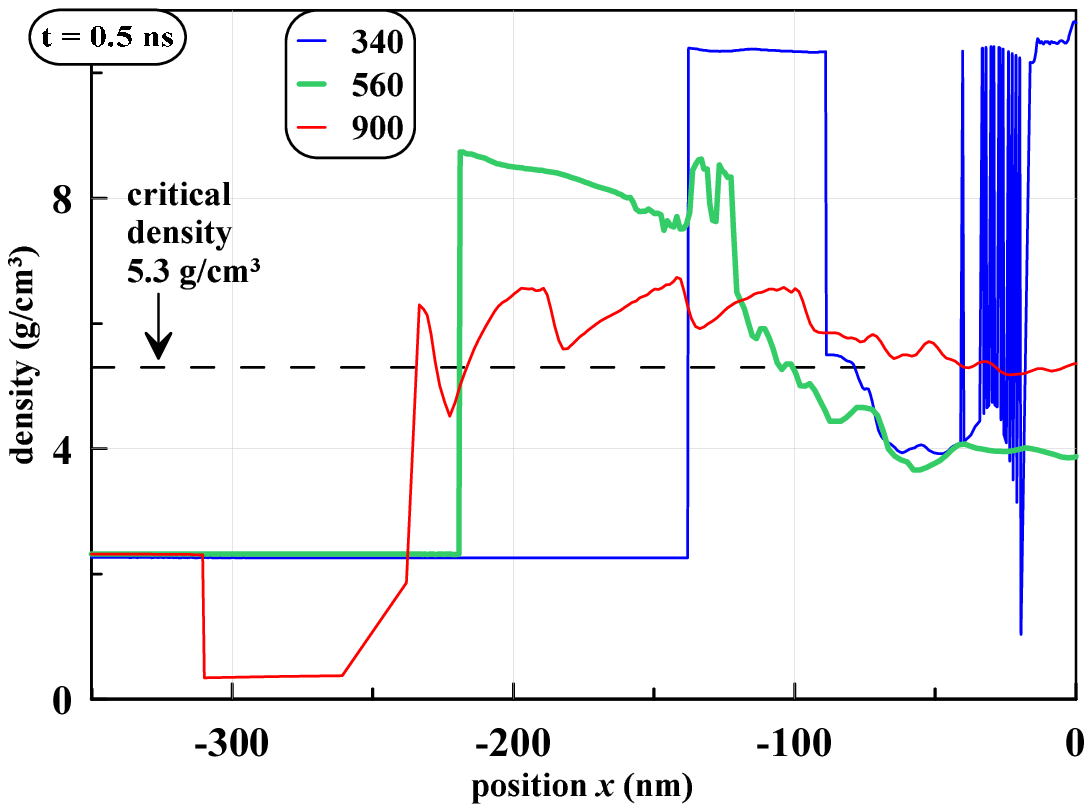}
\caption{\label{fig:15} Structure of the near CB layers.
The CB expands more in dielectric at the same instant of time as more energy $F_{abs}$ is absorbed.
Parameters of simulations are presented in Figures \ref{fig:12} and \ref{fig:13}.
We see that only in the strongest case $F_{abs}=900$ mJ/cm$\!^2$ the density of gold at the CB remains
below the critical density and thus at the gaseous side above the critical point at the $\rho,T$ phase diagram.
While in the two other cases these densities are at the liquid side of the diagram.
  }  \end{figure}


 Heated gold is cooled due to heat conduction into deeper layers and thanks to adiabatic cooling during expansion.
 Inertial confinement by rather dense dielectric limits a rate of expansion and thus the rate of cooling.
 It takes approximately 0.2 ns to cool the CB below critical temperature $T_{cr}$ in the case with $F_{abs}=560$ mJ/cm$\!^2.$
 In the case with $F_{abs}=900$ mJ/cm$\!^2$ the time interval for such cooling is longer, see Figures \ref{fig:13}-\ref{fig:15}.


 The temporal interval, when gold near a CB is in overcritical state, plays a special role in the process of nanoparticles (NP) formation.
 See discussion in the next Chapter.
 Here we have to discuss strong increase of temperature near the CB (see Figs. \ref{fig:12}, \ref{fig:13}) in the case of the most powerful action $(F_{abs}=900$ mJ/cm$\!^2)$
   from the sequence of the considered cases.
 It is obvious that temperature at the CB after absorption of a subnanosecond pulse $T_{CB}(t\approx \tau_L)$ increases as energy $F_{abs}$ increases while density $\rho_{CB}$ drops down.
 How high is temperature $T_{CB}(t\approx \tau_L)$ at the CB?
 From energy balance we have an estimate $T_{CB}(t\approx \tau_L)\sim F_{abs}/d/c,$ where $d\sim 2\sqrt{\chi_{Au}\tau_L},$ $c$ is heat capacity of gold per unit of volume.
 Our model of heat conduction \cite{INA.jetp:2018.LAL,Petrov.INA.KPM.jetp.lett:2013} describes dependence
  of thermal diffusivity $\chi_{CB}$ on density.
 Values of $\chi_{CB}$ and $c$ decreases as density decreases.
 Thus temperature $T_{CB}(t\approx \tau_L)$ isn't simply proportional to $F_{abs}$ as it is in the case with permanent values of $\chi_{CB}$ and $c.$


 Namely nonlinear dependence of a function $T_{CB}(t\approx \tau_L)$ on $F_{abs}$ causes appearance of the overheated contact layer in Figures \ref{fig:12} and \ref{fig:13}, see also Fig. \ref{fig:15}.
 In this layer temperature increases 5-7 times above the average level.
 While density drops 5-7 times below the average level, see Fig. \ref{fig:15}.
 Thus pressure is approximately the same across the matter of glass near the CB, in the overheated layer, and in the dense gold near this layer, see Fig. \ref{fig:14}.


 Transport properties of gold in the states with $\rho\sim 1$ g/cm$\!^3$ and $T\sim $ few eV are poorly known.
 Thus we cannot pretend that our conduction model \cite{INA.jetp:2018.LAL,Petrov.INA.KPM.jetp.lett:2013} is very accurate at these hot states of intermediate density
  between condensed and gaseous values.
 But we are sure that at least at the semi-quantitative level the model is true in these states.
 Therefore phenomenon with the nonlinear overheating have to take place but may be at slightly different energies $F_{abs}.$
 If we suppose absorption of gold at the level $A=0.3$ than incident intensity $I_{inc}\sim F_{abs}/A/\tau_L$ is $\sim 10^{11}$ W/cm$\!^2,$
    that is somewhere not far from the values corresponding to optical breakdown of dielectrics.
 This understanding seems limits achievable level of the highest temperatures $T_{CB}.$

\section{Disappearance of capillary barrier and strong enhancement of diffusion}

\begin{figure}       
   \centering   \includegraphics[width=0.75\columnwidth]{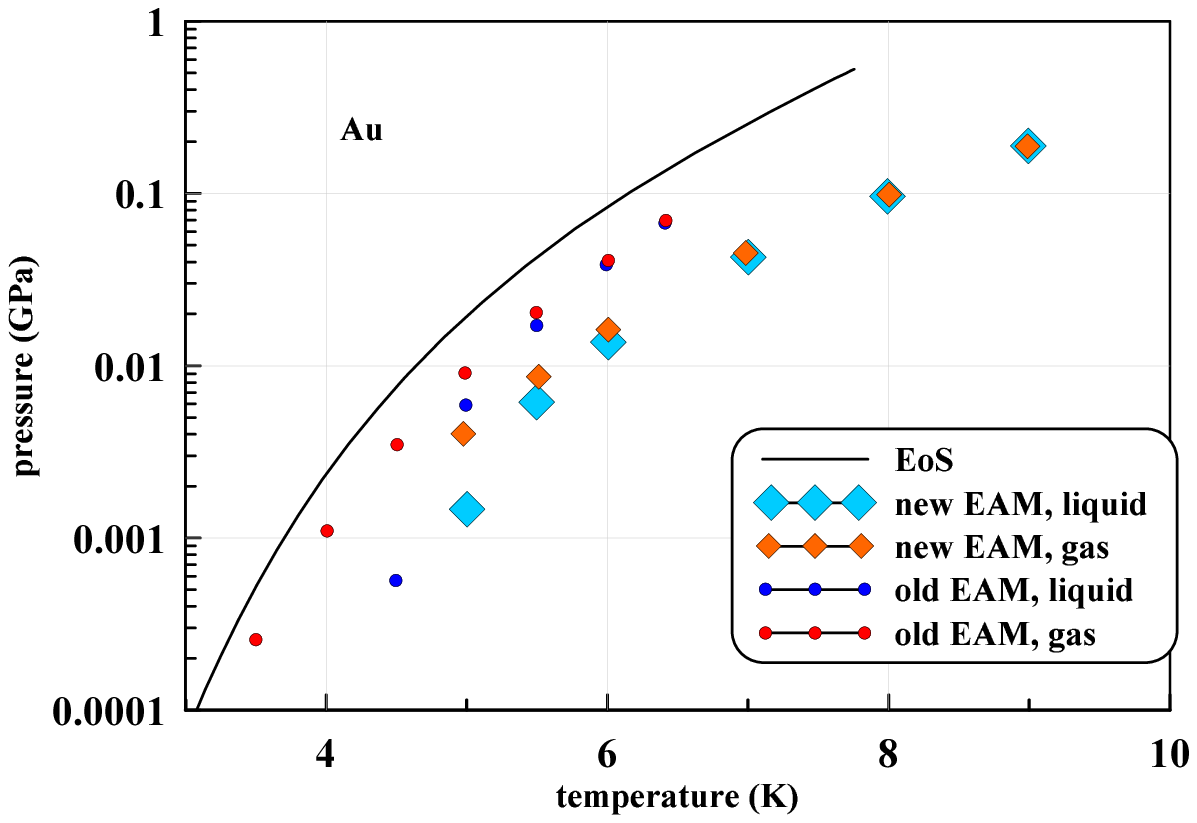}
\caption{\label{fig:16} Saturation pressure $p_{sat}(T)$ curves for Au from the wide-range equation-of-state (EoS) \cite{Bushman:1993,Khishchenko2002,Lomonosov_2007,rusbank1,rusbank2}
 and from MD/EAM simulations. The right end points of the curves correspond to the critical points.
 It is difficult to define $p_{sat}$ from MD/EAM at small pressures much smaller than 1 GPa.
 Simulation errors in the MD/EAM approach also increase near the critical point.
  }  \end{figure}

\begin{figure}       
   \centering   \includegraphics[width=0.7\columnwidth]{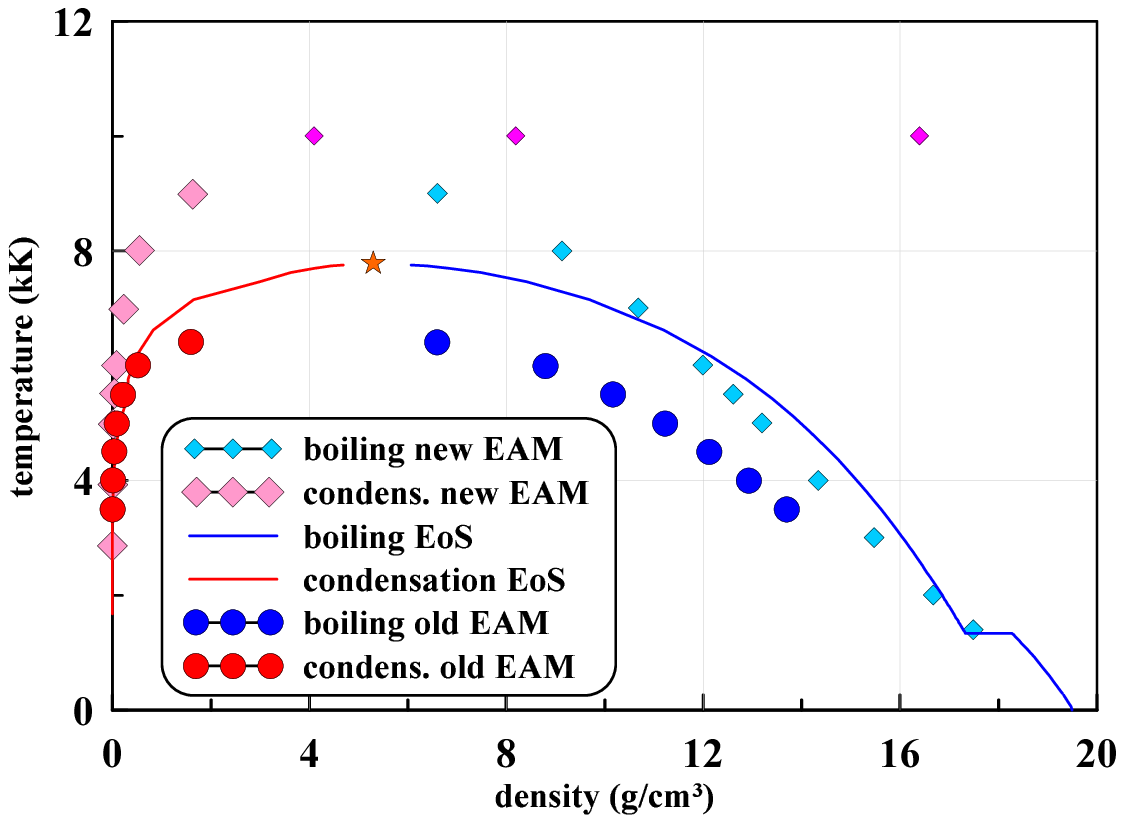}
\caption{\label{fig:17} The $\rho,T$ plane of a phase diagram for Au.
Comparison of the binodal curves from MD/EAM and EoS.
The 3 magenta rhombus shows places where diffusion coefficient in pure Au and Au-wt mixture was evaluated, see Table 1.
MD/EAM reproduces accurately melting temperature of Au.
  }  \end{figure}

\begin{figure}       
   \centering   \includegraphics[width=0.7\columnwidth]{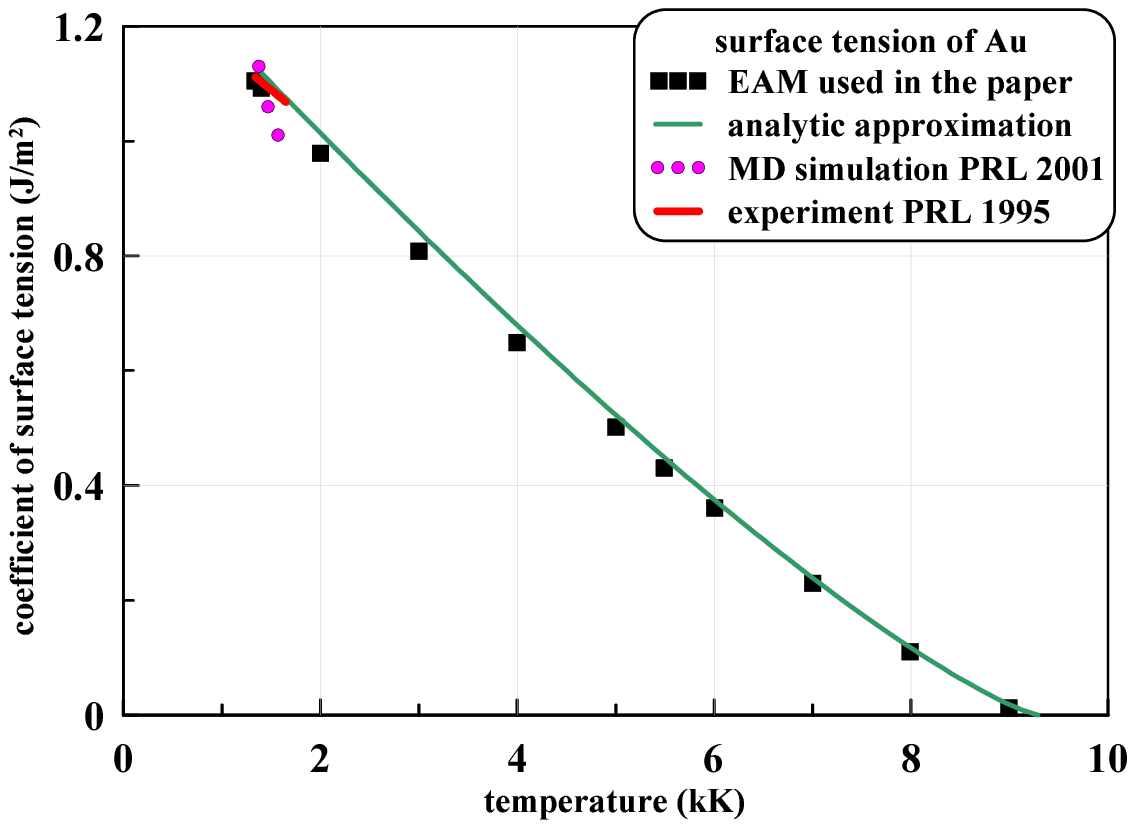}
\caption{\label{fig:18} Behavior of surface tension of gold $\sigma(T)$ along the boiling curve
 shown in Figures \ref{fig:16} and \ref{fig:17}.
 Black squares relate to the MD/EAM calculations done using the new EAM potential.
 Calculation of $\sigma(T)$ by the old EAM is presented in Fig. 15 in paper \cite{Inogamov2015nanoBump}.
 The old EAM gives approximately twice lower values of $\sigma(T).$
 See discussion in \cite{Zhakhovsky-sigma:1999} about evaporation and surface tension.
 Green continues curve gives approximation $\sigma=\sigma(T_3)((1-T/T_{cr})/(1-T_3/T_{cr}))^{5/4}$
  from \cite{Semenchenko:sigma-T:1961}, see also \cite{Inogamov2015nanoBump}.
 Here $T_3$ and $T_{cr}$ are temperatures in the triple and critical points.
 The PRL1995 and 2001 relates to the data from \cite{PRLsigma:1995,PRLsigma:2001}.
 They cover only short temperature intervals.
   }  \end{figure}


 As follows from Figures \ref{fig:01}, \ref{fig:12}-\ref{fig:15} the CB is above the critical point for gold for some temporal interval.
 Critical point of water (220 bar, 647 K, 0.3068 g/cm$\!^3)$ is much lower than the critical point of gold.
 Thus both liquids are in the overcritical states.
 Then they behave like the non-ideal gases.
 Surface tension at the contact boundary (CB) between them becomes zero.
 While diffusion coefficient increases.
 In condensed matter the diffusion process is slow because an atom loses time doing many oscillations inside its potential well
   before it tranships to the neighbor well.




 We use molecular dynamics (MD) simulation with EAM (embedded atom method) interatomic potential for gold
  to calculate saturation pressure (Fig. \ref{fig:16}), the boiling and condensation curves (Fig. \ref{fig:17}),
    and surface tension at the boiling curve (Fig. \ref{fig:18}).
 We modify our previous EAM potential for Au-Au interaction described in \cite{Zhakhovskii.EAM:2009}.
 The modified potential provides the correct dependence of surface tension coefficient on temperature
   shown in Fig. \ref{fig:18} and dimer Au-Au binding energy.
 Mechanically both potentials are close to each other (equilibrium density, compressibility, Hugoniot adiabat).
 Melting temperatures are very near to the experimental value for both potentials.
 We refer to these EAM potentials as old and new EAM potentials in Figures \ref{fig:16} and \ref{fig:17} and below in the text.


 Figure \ref{fig:16} presents the binodal curve at the $T,p$ phase plane.
 We see that both EAM potentials underestimate saturation pressure at high temperatures near the critical point
  relative to the data from \cite{Bushman:1993,Khishchenko2002,Lomonosov_2007,rusbank1,rusbank2}.
 While critical temperature is or higher for the new EAM potential, or lower for the old one.
 It should be said that the exact value of critical pressure and temperature of gold still remains unknown.
 Thus in reality we cannot definitely conclude what set of data is more close to the real gold.


 Achievement of the thermodynamic equilibrium between liquid and its vapor in the MD run becomes more difficult as temperature and thus pressure decrease
   (because the rate of evaporation drops down sharply thus simulation time increases).
 But the equilibrium is achieved in all cases presented.
 The lowest pressures are of the order of 1 bar, see Fig. \ref{fig:16}.
 The difference between gas and liquid pressures in Fig. \ref{fig:16} for the same EAM potential follows from large errors in definition of pressure in liquid when pressure is low.
 Indeed, in liquid pressure is a sum of two large values having opposite signs: thermal and virial pressures.
 Unfortunately these errors leave rather large systematic shift.
 Therefore pressures in liquid in Fig. \ref{fig:16} are below the gaseous pressures.
 We have to use the gaseous pressures as saturation pressure for a given EAM potential.
 It is interesting that the pair interatomic potentials are less sensitive to this problem than the EAM (embedded atom method) potentials (they include many-body interactions).
 Liquid pressures are presented to familiarize readers with quality of results and real problems existing in MD approach.


 Fig. \ref{fig:17} shows boiling and condensation curves at the density-temperature plane.
 Equation-od-state (EoS) \cite{Bushman:1993,Khishchenko2002,Lomonosov_2007,rusbank1,rusbank2}, and two EAM potentials are compared.
 We see that at the $\rho,T$ plane the new and old potentials limit the EoS data from above and from below.


 Molecular dynamics (MD) simulation of atom diffusion in the hot binary mixture Au-water
  was performed with using three different potentials describing interactions between atoms
   Au-Au, water-water, and Au-water.
  The new EAM potential was used for describing of Au-Au interactions.
 Water is considered in our simulation as a monatomic system,
  for which the EAM potential was developed in \cite{INA.jetp:2018.LAL},
   see the data files on the project  
   \url{https://www.researchgate.net/project/Development-of-interatomic-EAM-potentials.}
  It provides the correct mechanical properties of liquid water including the density and sound speed in ambient conditions,
  and shock Hugoniot as well.

 
\begin{figure}       
   \centering   \includegraphics[width=0.7\columnwidth]{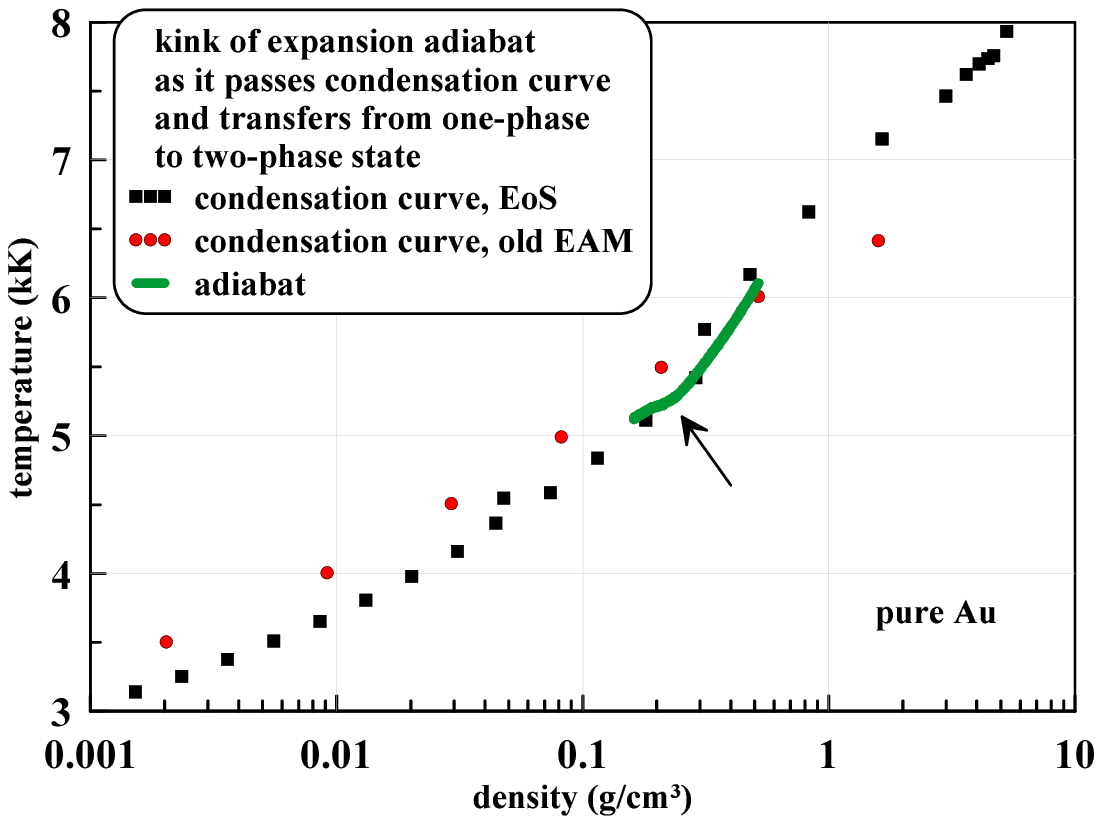}
\caption{\label{fig:19} Wide range of a phase diagram showing the small piece of the adiabatic curve near the point where the curve passes through the binodal.
 The EoS and the old EAM binodals are shown.
 They are close to each other near this particular intersection points.
 MD/EAM simulation are done using the old EAM potential.
   }  \end{figure}

 Due to lack of information, the interaction of Au atom and water molecule is greatly simplified
  to a pairwise potential between Au and ``water'' atoms using the van der Waals radii
   0.166 nm for Au and 0.152 nm for oxygen atom, and a strong repulsion $\sim r^{-11}$ is assumed.
 The usage of larger van der Waals radius 0.19 nm of water for our ``water'' atom
  reduces the diffusion coefficient of Au atoms in the simulated 50/50 binary mixture by about 10-20\%;
   50/50 is the ratio of the numbers of atoms per unit of volume;
     partial densities are 4.1 and 0.4 g/cm$\!^3$ for Au and water, respectively.
 Results of MD simulations relating to calculations of diffusion coefficient $D$ in the overcritical states of gold
   and gold-water mixture are presented in Table 1.
 Corresponding MD simulations were done using new EAM potential.
 Positions of corresponding points (where coefficient $D$ was calculated) are marked as magenta rhombus in Fig. \ref{fig:17}.




\begin{table}[hbt]
\caption{
Self-diffusion Au-Au coefficient $D_{Au-Au}$ in homogeneous hot temperature gold $T=10$ kK
   in the overcritical states shown in Fig. \ref{fig:17} by the magenta rhombus.
 The last line gives the diffusion coefficient $D_{Au-wt}$ of Au atom in the homogeneous Au-water mixture $T=10$ kK
   with atomic partial concentrations 50/50
     and partial densities 4.1 and 0.4 g/cm$\!^3$ of Au and water, respectively. We use new EAM in calculations.}
\centering
\begin{tabular}{|c|c|c|c|}
\hline
$\rho$ of Au                    [g/cm$\!^3$]                   & 4.1   & 8.2    & 16.4      \\
\hline
$D_{Au-Au}$              $[\times 10^{-3}$ cm$\!^2/$s]  & 1.47  & 1      & 0.41      \\
\hline
$D_{Au-wt}$ $[\times 10^{-3}$ cm$\!^2/$s]  & 1.13  &        &           \\
\hline
\end{tabular}
\end{table}

\begin{figure}       
   \centering   \includegraphics[width=0.75\columnwidth]{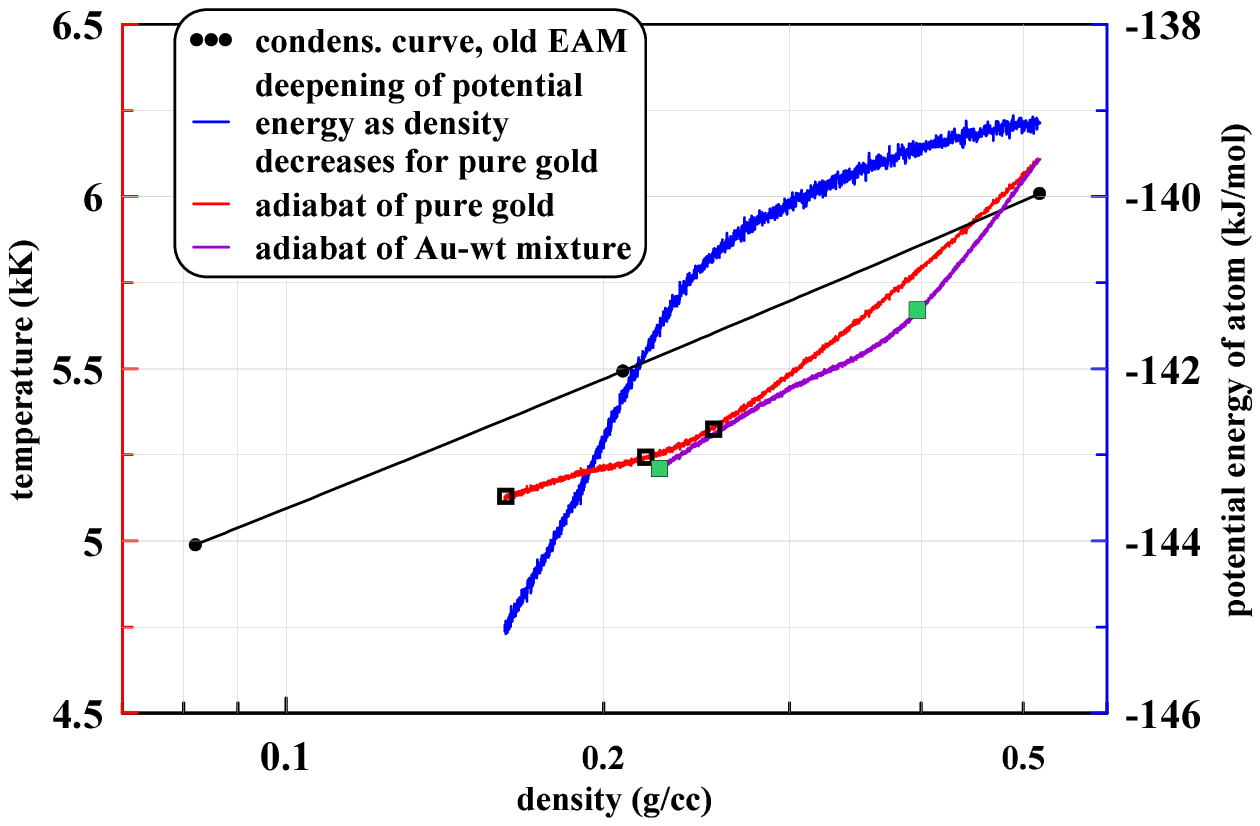}
\caption{\label{fig:20} The detailed view of the intersection region taken from previous Figure \ref{fig:19}.
 Two adiabatic curves are presented. One corresponds to pure Au, while another one - to 50/50 mixture of Au and water (in numbers of atoms per unit of volume).
 The axis of density gives a sum of Au and water densities in the case of a mixture.
 The condensation curve relates to the case of pure Au, see also Figures \ref{fig:17} and \ref{fig:19}.
 It is difficult to find position of the gold condensation curve in case of mixture.
 It should differ from the binodal for pure gold.
 Approximately the condensation curve for the 50/50 mixture is located slightly above the point where the 50/50 adiabatic curve has the kink.
 Snapshots of the process is demonstrated in Figures \ref{fig:21} and \ref{fig:22} below.
   }  \end{figure}

\section{Cooling, condensation, and formation of nanoparticles}    \label{sec:06:condens=19.20-21.22}


 There is significant gold-water atomic inter-diffusion mixing process
   when matter near the CB exist in the overcritical states; see Figures \ref{fig:12}-\ref{fig:15} about duration
     of residence in the overcritical states and Fig. \ref{fig:17} with Table 1 about the enhanced value of
      diffusion coefficient $D$ in these states.
 Thus the jump like CB diffuses into a smearing layer.
 Thickness of this layer is $l_{diff} \sim 2\sqrt{D\, t}.$
 Taking durations $t$ from few to few tens of nanoseconds we obtain values of $l_{diff}$ from few tens to $\sim 10^2$ nm.


 Hydrodynamic expansion causes cooling of mixture thanks to work done for expansion.
 Gradually temperature drops below the critical value for gold.
 Then condensation into clusters of Au atoms begins.
 These clusters are in Brownian movement, sometimes meet each other and stick together.
 Atoms surrounding a cluster condensate to a cluster and evaporate from it,
   thus mass and energy exchange between atoms and a cluster takes place.
 Our MD simulations presented below show that temperatures of atoms and clusters are approximately the same.
 Thus collisional removal of the positive energy excess from growing in size (and hence heating) cluster is fast enough to support equality of temperatures.


 The hydrodynamic expansion rate is defined by the gradient of velocity $dv/dx$ near the CB.
 At the nanosecond time scale this gradient is of the order of $dv/dx|_{CB}\sim 10^8$ 1/s,
   see Figures \ref{fig:02}, \ref{fig:04}, and \ref{fig:06}.
 Therefore expansion velocity of the CB diffuse layer $dv/dx|_{CB}*l_{diff}$ is a few meters per second.


 Following these conditions we run several MD simulations to estimate temporal range
  of nucleation and growth of nanoparticles (NP) as pure gold or the 50/50 mixture (in concentrations of Au and water)
    intersect a condensation curve shown in Fig. \ref{fig:17}.
 Calculations are done using the old EAM potential.
 Condensation starts with some delay in time after the intersection.
 Of course, this delay depends on the rate of expansion - in the case of very slow expansion the condensation begins very close to the intersection point.

\begin{figure}       
   \centering   \includegraphics[width=\columnwidth]{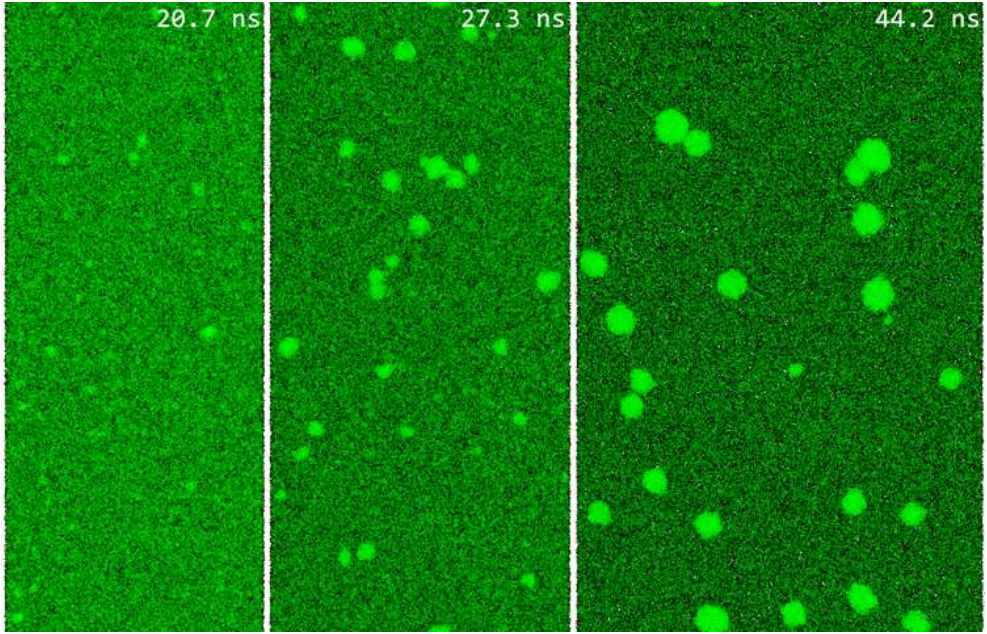}
\caption{\label{fig:21} The course of condensation process during expansion of initially gaseous Au.
 Appearance of the viable clusters begins not immediately as a system of Au atoms crosses an equilibrium curve, see Fig. \ref{fig:20}.
 Let's mention that unstable (appearing/disappearing) clusters may exist even in gaseous state.
 They contribute to decrease of pressure relative to pressure of ideal gas with same temperature and atom number density.
 The three frames shown correspond to the instants listed in Table 2 and marked by empty squares in Fig. \ref{fig:20}.
 The $x,y$ plane is demonstrated. Horizontal direction of expansion is $x.$ Its expansion velocity is 2 m/s. Vertical size along $y$-axis is 200 nm.
   }  \end{figure}

\begin{figure}       
   \centering   \includegraphics[width=\columnwidth]{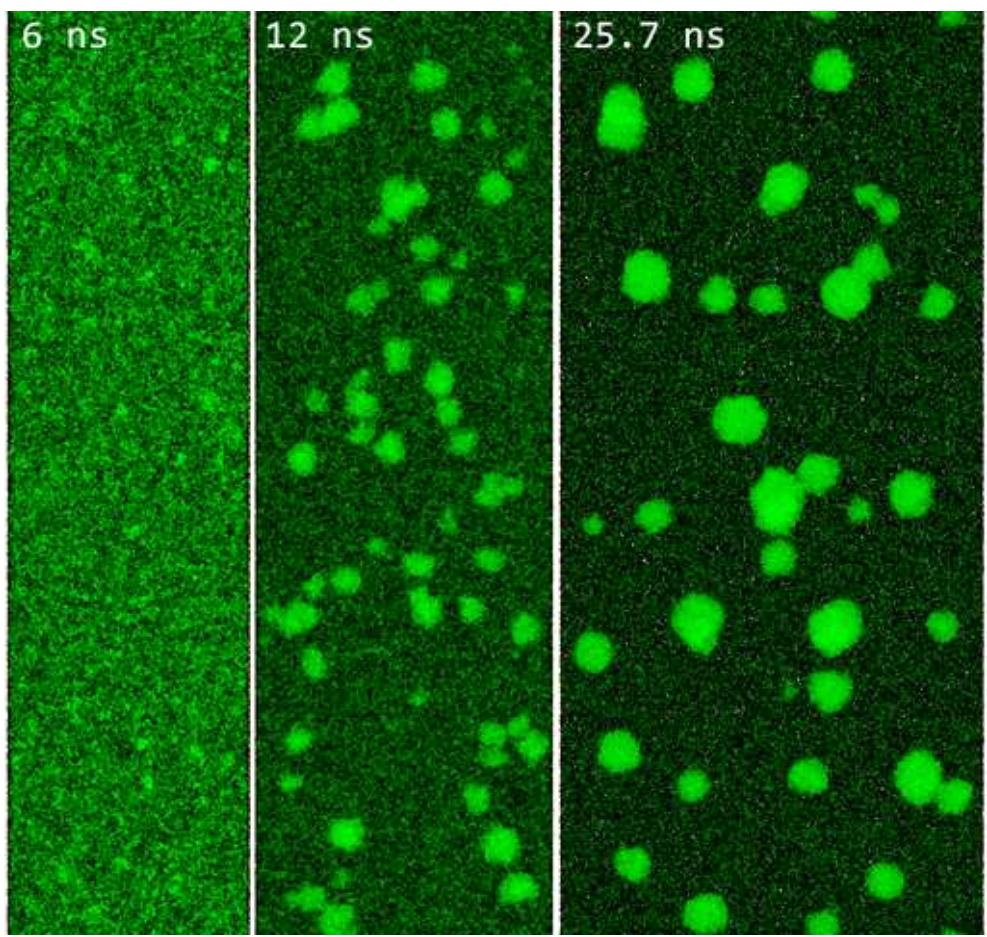}
\caption{\label{fig:22} Condensation into gold nanoparticles as Au-water mixture at the $\rho,T$ phase plane decreases below the equilibrium curve for gold in mixture.
The left and right frames relate to the two green squares in Fig. \ref{fig:20}, see also Table 3.
We add the intermediate frame $t=12$ ns to better show the process of nanoparticles (NPs) formation and growth.
Sizes of NPs may be estimated by direct measuring, compare with the vertical dimension (this is the $y$ axis) of the computational box equal to 200 nm.
Adiabatic expansion proceeds along the horizontal direction (this is the $x$ axis) expanding with velocity 2 m/s.
   }  \end{figure}





 Figures \ref{fig:19} and \ref{fig:20} show how the adiabatic curves kink as they pass into the two-phase region.
 The kink corresponds to the transition from the fast temperature drop with expansion to the slow temperature drop.
 The fast drop (rather large effective adiabatic index $\gamma)$ relates to the one-pase behavior (Au is in a gas state).
 The slow drop (decrease of $\gamma)$ begins when liquid component appears in a two-phase vapor-liquid mixture.
 Release of evaporation energy through condensation during volume expansion reduces the rate of temperature decrease.





 Fig. \ref{fig:19} presents global view of the EoS and old EAM boiling curves and the MD simulated adiabatic curve crossing the boiling curve.
 The MD simulation is intended for definition of the transition piece from the one-phase to the two-phase behavior
  along the adiabatic expansion process
    for our particular conditions: velocity gradient $\sim 10^8$ 1/s and the diffusion thickness of the order of few tens of nm.
 We are interested to know how long this transition lasts and how large nanoparticles are at the end of our temporal range.
 The MD simulation is rather difficult because duration is long (tens of ns) and system should consists from large number of atoms $\sim 10^6.$
 Therefore we limit our self to a narrow time interval just before the intersection
  (omitting trivial one-phase piece of the gaseous adiabatic curve)
    and we follow the process during a time scale typical for our problem.
 This explains why the piece of the adiabatic curve is short in Fig. \ref{fig:19} relative to the field shown.


\begin{table}
\caption{
The lines in the Table present the list of the points at the adiabatic curve for pure Au in Fig. \ref{fig:19} and \ref{fig:20}.
 In Fig. \ref{fig:20} three last of these points (the three bottom lines here) are marked by the empty squares.
 These three points correspond to the digits 20.7, 27.3, and 44.2 ns at the maps of potential energy shown in Fig. \ref{fig:21}.}
\centering
\begin{tabular}{|c|c|c|c|}
\hline
instant [ns]   & $T$ [kK]   & $p$ [bar]    & $\rho$ [g/cm$\!^3$]      \\
\hline
0              & 6.11       & 436          & 0.518                    \\
\hline
20.7           & 5.32       & 194          & 0.254                    \\
\hline
27.3           & 5.24       & 164          & 0.219                    \\
\hline
44.2           & 5.13       & 126          & 0.162                    \\
\hline
\end{tabular}
\end{table}


\begin{table}
\caption{
The three lines in the Table present the list of the points at the adiabatic curve for Au-water mixture in Fig. \ref{fig:20}.
 In Fig. \ref{fig:20} the two last of these points (the two bottom lines) are marked by the filled green squares.
 These points correspond to the digits 6 and 25.7 ns at the maps of potential energy shown in Fig. \ref{fig:22}.
 Here the last two columns give total pressure and total density of a mixture (a sum of Au and water contributions).}
 \centering
\begin{tabular}{|c|c|c|c|}
\hline
instant [ns]   & $T$ [kK]   & $p$ [bar]    & $\rho$ [g/cm$\!^3$]      \\
\hline
0              & 6.10       & 936          & 0.516                    \\
\hline
6              & 5.67       & 643          & 0.397                    \\
\hline
25.7           & 5.21       & 324          & 0.226                    \\
\hline
\end{tabular}
\end{table}



 Fig. \ref{fig:20} demonstrates shape of the transient kinks for pure Au and for a mixture Au-wt.
 This is the local vicinity of the points where the adiabatic curves for pure and mixed Au cross the curves of thermodynamic equilibrium for pure and mixed Au,
   compare with Fig. \ref{fig:19}.
 MD simulations with old EAM interatomic potential were used to obtain these two adiabatic curves.
 The simulation boxes for the both cases have fixed in time periodic boundary conditions at the lateral walls, that is in the $y,z$ directions.
 Lateral dimensions $y\times z$ are $200\times 100$ nm.
 The moving walls are imposed in the $x$ direction.
 There are reflecting potentials connected with these $x$-walls.
 They return the Au or Au and water atoms colliding the $x$-wall back into box.
 Reflection from the expanding walls extracts small (velocity of expansion is much less than thermal velocity) part of kinetic energy of colliding atoms
   thus adiabatically cooling a whole system in the box.


 Initial densities and temperatures are given in Fig. \ref{fig:20} and Tables 2 and 3.
 Initial distance between the $x$-walls is $\Delta x_0 = 40$ nm.
 This starting distance is the same in pure Au and the mixture cases.
 Expansion velocity $v_x|_{expan} = d\Delta x/dt$ is 2 m/s.
 It is also the same in the both case.
 Distance between the $x$-walls grows with time and density decreases $\rho=\rho_0 \Delta x_0/(\Delta x_0 + v_x|_{expan}\, t).$
 This is a model of spatial expansion of diffusion mixture in real conditions during expansion and cooling.


 Figures \ref{fig:21} and \ref{fig:22} present pictures of evolutions of pure Au (Fig. \ref{fig:21}) and mixed systems, respectively.
 Detailed description of simulation approach and obtained results is given in the paper \cite{Zhakhovsky-ANGEL:2018} presented at the ANGEL-2018 conference.
 Movies of evolution are attached to this paper.
 Figures \ref{fig:21} and \ref{fig:22} show how nanoparticles nucleate and grow.
 They corresponds to the view along the $z$-axis.
 The simulation box is 100 nm long in the $z$-direction.
 Figures \ref{fig:21} and \ref{fig:22} are composed from the $x,y$ set of the individual pixels.
 Every pixel contains information about potential energy averaged along the $z$-direction.


 More light green pixel marks position where potential energy is deeper.
 The light green spots are the nanoparticles (NP).
 Their sizes are of the order of ten nanometers at the final stages in both cases of pure Au (Fig. \ref{fig:21}) or Au-water mixture \ref{fig:22}).
 Atoms sticking together into NP have profit in their interaction energy.
 Sticking of part of Au atoms in the two-phase system does deeper potential energy of the whole system (averaging along $x,y,z$-directions).
 This is shown in Fig. \ref{fig:20} by the dependence relating to potential energy.
 Figures \ref{fig:21} and \ref{fig:22} show the same but with averaging only along the $z$-direction.
 Temperatures in Figures \ref{fig:21} and \ref{fig:22} are higher than melting temperature of gold.
 Thus NP remains in liquid states in the temporal range considered.







\begin{figure}       
   \centering   \includegraphics[width=0.75\columnwidth]{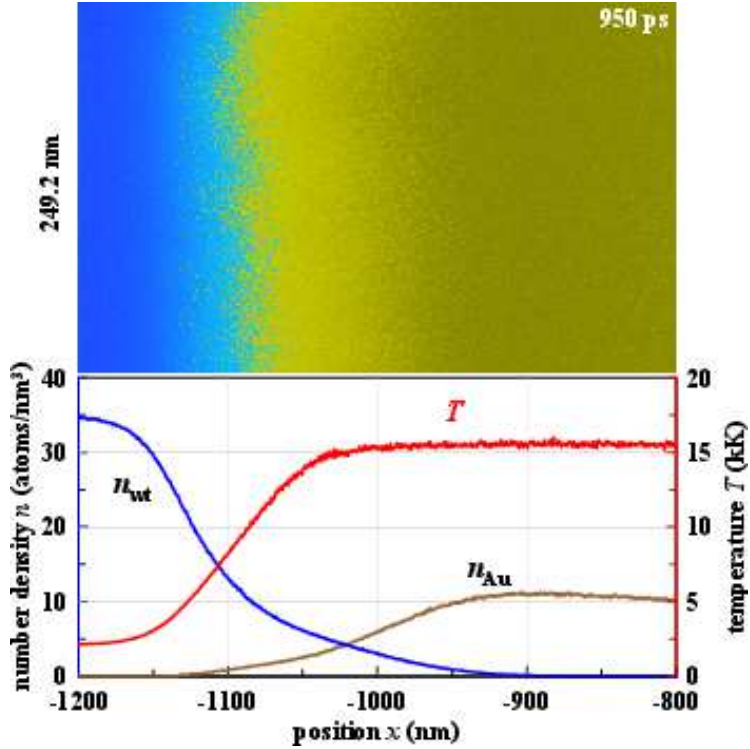}
\caption{\label{fig:23} The upper panel presents the snapshot of atomic number density map of Au (yellow colors)
 moving to the left side with velocity $\approx 700$ m/s at this instant; degrees of yellow are proportional to local atomic density presented in the bottom panel.
Au displaces water which resists to displacement and thus causes deceleration of Au.
Deceleration and density difference are the reasons that result in appearance of Rayleigh-Taylor instability (RTI) evidently seen \cite{LZ-bulk-LAL:2017,LZ+Stephan:2018.LAL,INA.jetp:2018.LAL}
 at the lower values of $F_{abs}.$
Surface tension imposes clear capillary spatial scale for the growing perturbations \cite{LZ-bulk-LAL:2017,LZ+Stephan:2018.LAL,INA.jetp:2018.LAL}.
But here at enhanced laser action the diffusion is stronger and it suppresses development of RTI while the capillary scale shrinks to zero.
The bottom panel gives the number densities instant profiles of Au and water atoms corresponding to the instant shown in the upper panel.
The map from the upper panel is integrated in the lateral dimensions for Au and water separately to obtain these profiles.
   }  \end{figure}

\section{Molecular dynamics simulation of ablation in water}      \label{sec:07}










 From the sequence of three shots compared in Figures \ref{fig:09}, \ref{fig:09} and \ref{fig:12}-\ref{fig:15}
    we see that increase of absorbed energy $F_{abs}$ causes serious consequences.
 The main feature is that with this increase a metal near a CB transfers from undercritical to overcritical conditions and remains their for rather long time.
 Undercritical (condensed) and overcritical (uncondensed, gaseous) conditions differ qualitatively.
 In the undercritical case (relatively small $F_{abs})$ the liquid gold contacts with liquid; our fluences are above melting threshold for Au.
 Thus to penetrate into liquid and to form NPs the Au atoms have to overcome surface barrier through evaporation/dissolution (the first step)
   and after that they diffusively drift out from a CB (the second step).
 At the last step the expansion and cooling lead to condensation and formation of NPs, see Figures \ref{fig:20} and \ref{fig:22}.


 In the overcritical states the first step disappears together with surface tension.
 This is very important.
 Indeed, low values of saturated pressures of evaporated/dissolved Au atoms outside the boundary of a continuous condensed phase (outside the CB)
  strongly limits amount of Au atoms in a diffusion layer and thus limits amount of NPs.
 In the overcritical states only the second and third steps are present.
 This enhances diffusion - now diffusion starts not from evaporated/dissolved atomic number density of Au near the CB
   but from much larger value of atomic number density directly in the continuous phase.


 We perform a large scale MD simulation to follow ablation of gold into water at absorbed energy $F_{abs}$ larger than in the case previously considered in paper \cite{INA.jetp:2018.LAL}.
 Comparison of these two cases together with detailed description of evolution of the flow is given in the paper \cite{Zhakhovsky-ANGEL:2018} presented at the ANGEL 2018 conference.
 Here in Fig. \ref{fig:23} rather late instant from MD simulation with large $F_{abs}$ is presented.
 We see that Au-water interpenetration significantly overcomes the estimate $l_{diff}=2\sqrt{D\, t}=20$ nm for $D=10^{-3}$ cm$\!^2/$s and $t=1$ ns.
 May be enhancement of interpenetration is due to presence of some embryos of Rayleigh-Taylor instability.
 Indeed, it is known that diffusion is a laminar like process which produces very flat atomically mixed layer; diffusion smooths away lateral roughness.
 But in Fig. \ref{fig:23} we see significant inhomogeneities in the direction transverse to the direction of longitudinal expansion.

\section{Conclusion}

 Above ablation of gold into transparent media is analyzed.
 Initial stages of formation of nanoparticles (NP) are considered.
 These stages relate to tens and hundreds of nanoseconds (see Section 6) well before bubble dynamics comes into play.
 All thermodynamical (with first order phase transitions) and hydrodynamical aspects have to be included to come to understanding of appearance of primary NPs.
 It is found that

 \noindent * Deceleration of the contact boundary (CB) at the nanosecond time scale is approximately exponential
 $v_{CB}\propto \exp(-t/t_{dec})$ with the e-folding time $t_{dec}$ of the order of 0.5-1 ns.
 This law describes deceleration in the 1-0.1 km/s range of velocities $v_{CB}$ after ultrashort and subnanosecond pulses.


 \noindent * Maximum shift of the CB from its initial position is rather limited and is of the order of micron.


 \noindent * Temperature $T_{CB}$ of gold at the CB nonlinearly depends on absorbed energy $F_{abs}.$
 Temperature $T_{CB}$ strongly increases for energies $F_{abs}\sim 1$ J/cm$\!^2$ up to values $\sim 10^5$ K for subnanosecond pulses.
 These values are much above critical temperature.
 While density of gold $\rho_{CB}$ decreases down to values $\sim 1$ g/cm$\!^3.$


 \noindent * Ablation motion triggered by a subnanosecond pulse is order of magnitude more stable against Rayleigh-Taylor instability (RTI) in comparison with an ultrashort action,
     see Fig. \ref{fig:11}.
  This is simply caused by the shortening of deceleration stage because more than a half of the heating pulse is spent to accelerate the CB.


 \noindent * Another important reason to limit the RTI perturbations growth is connected with high temperatures,
   decrease of the density ratio at the CB, and enhancement of diffusion.
 Enhanced diffusion is a powerful stabilizer of RTI.
 Diffusive smearing grows as $l\propto t^{1/2}$ while for RTI in the case with permanent deceleration the law of growth is $l\propto t^2.$
 Thus diffusion always dominates at the early stages while the RTI wins this competition later in time.
 But in laser ablation the deceleration quickly decays in time.
 Thus there is only limited interval of times when perturbations can develop thanks to RTI.
 In this situation the diffusion process suppresses RTI.
 Then RTI is removed from the process of fabrication of NPs at the capillary scale.
 Appearance of NPs of the capillary scale was observed in simulations \cite{LZ-bulk-LAL:2017,LZ+Stephan:2018.LAL,INA.arxiv:2018.LAL,INA.jetp:2018.LAL}.




\section*{References}
\bibliographystyle{elsarticle-num}
\bibliography{LAL}
\end{document}